\documentclass[a4paper,12pt]{article}
\usepackage[utf8]{inputenc}
\usepackage[english]{babel}
\usepackage{hyperref}
\usepackage[text={6.5in,9in},centering,a4paper]{geometry}
\usepackage{amsmath,amssymb}
\usepackage{bbold}
\usepackage{graphicx}
\graphicspath{{./}{Figs/}}
\title{The benefits of peer transparency in safe workplace operation post pandemic lockdown.}
\author{Arkady Wey$^{(1)}$, Alan Champneys$^{(2)}$, Rosemary J. Dyson$^{(3)}$,\\ Nisreen A. Alwan$^{(4),(5),(6)}$ and Mary Barker$^{(5),(7)}$}

\date{6th July 2020}

\usepackage{graphicx}

\begin{document}

\maketitle
\noindent {\small $(1)$ Industrially Focused Mathematical Modelling (InFoMM) EPSRC Centre for Doctoral Training, Mathematical Institute, University of Oxford, Oxford OX2 6GG, UK;}\\ {\small $(2)$ Department of Engineering Mathematics, University of Bristol, Bristol BS8 1TR, UK;}\\ 
{\small $(3)$ School of Mathematics, University of Birmingham,  Birmingham  B15 2TT, UK;}\\
{\small $(4)$ School of Primary Care, Population Sciences and Medical Education, University of Southampton, Southampton SO16 6YD, UK;}\\ 
{\small $(5)$ NIHR Southampton Biomedical Research Centre, University of Southampton and  University Hospital Southampton NHS Foundation Trust, Southampton SO16 6YD, UK; }\\
{\small $(6)$ NIHR Applied Research Collaboration (ARC) Wessex, Southampton, UK;}\\
{\small $(7)$ MRC Lifecourse Epidemiology Unit, University of Southampton  and University Hospital Southampton NHS Foundation Trust, Southampton SO16 6YD, UK.
}
\begin{abstract}
The benefits, both in terms of productivity and public health, are investigated for different levels of engagement with the test, trace and isolate procedures in the context of a pandemic in which there is little or no herd immunity.  Simple mathematical modelling is used in the context of a single, relatively closed workplace such as a factory or back-office where, in normal operation, each worker has lengthy interactions with a fixed set of colleagues. 

A discrete-time SEIR model on a fixed interaction graph is simulated with parameters that are motivated by the recent COVID-19 pandemic 
in the UK during a post-peak phase, including a small risk of viral infection from outside the working environment. Two kinds of worker are assumed,  
\emph{transparents} who regularly test, share their
results with colleagues and isolate as soon as a contact tests positive for
the disease, and \emph{opaques} who do none of these.
Moreover, the simulations are constructed as a ``playable model" in which the transparency level, disease parameters and mean interaction degree can be varied by the user. The model is analysed in the continuum limit. 

All simulations point to the double benefit of transparency in maximising productivity and minimising overall infection rates. 
Based on these findings, public policy implications 
are discussed on how to incentivise this mutually beneficial behaviour in different kinds of workplace, and simple recommendations are made.
\end{abstract}

\section{Introduction}

This study is inspired by the current UK situation (June 2020) as the
nation attempts to restart the economy in the aftermath of the
COVID-19 virus infection peak. The general structure of our
mathematical model and the parameter values chosen are specific to
that case study.  The results are nevertheless intended to be
applicable to more general situations in any modern society where
there is a residual risk of infection from a virus or other pathogen
with insufficient natural immunity in the general population.

There has been much recent evidence to suggest that the most effective
containment measure in a human epidemic with relatively small
proportions of infectious individuals is that of rapid testing,
contact tracing and isolation of those in the contact group
\cite{Kucharski,DELVE}. The effectiveness of such a strategy is
thought to be a function of the basic reproduction number of the
infection, known as the $R$-value; this gives, on average, the number
of new infections that each infection generates.  However, $R$ itself
is a crude measure, as its instantaneous value will be a function not
just on the basic disease dynamics, but on the behaviour of infectious
individuals.

The seminal paper of Keeling and Eames \cite{Keeling} introduced ideas
from graph theory to epidemiology, where the nature of interactions
between infected and susceptible individuals defines a dynamic contact
network. Ideas from modern network science, such as the degree
distribution, can then be used to estimate statistical properties of
the infection, such as the $R$-number, and to evaluate the
effectiveness of different potential treatment strategies; see
\cite{Pastor} for a relatively recent review of the state of the art.

The majority of studies that have looked at contact tracing as an
effective means of viral control have considered the question at a
general population level. We note the recently published studies
\cite{Kucharski,DELVE} which model the requirements of an effective
testing, tracing and isolation strategy to avoid a second-wave of the
COVID-19. One should not however under-estimate the required effort.
For example, based on data obtained in a unique collaboration with
BBC, Kuchraski {\em et al} conclude \emph{`in a scenario where there
  were 1,000 new symptomatic cases that met the definition to trigger
  contact tracing per day [$\ldots$] 15,000--40,000 contacts would be
  newly quarantined each day. [$\ldots$] A high proportion of cases
  would need to self-isolate and a high proportion of their contacts
  to be successfully traced to ensure an effective reproduction number
  that is below one in the absence of other measures.'} This finding
is even more stark when combined with results, e.g.~\cite{Periera},
that suggest that social isolation needs to happen sufficiently
quickly to be effective.

The most important features of any public health campaign built around
testing, contact tracing and isolation is the degree of compliance in
the general population, which can vary with the method used
\cite{Read}. Evidence presented to Scientific Pandemic Influenza group
on Behaviour (SPI-B) as part of the advice offered to the UK
Government Scientific Advisory Group for Emergencies (SAGE) indicated
a shortlist of factors that might help to promote compliance with and,
adherence to, all behaviours that minimise transmission of SARS-CoV-2
infection. These apply equally to compliance with testing and contact
tracing. Factors included messaging that increased perceptions of
risk, clear communications from Government identifying what behaviours
the public should adopt, encouraging support from the community so
creating social norms for infection-limiting behaviours and
importantly, actions making it as easy as possible for people testing
positive to isolate \cite{SPI-B3,SPI-B1,SPI-B2}. Recent studies have
indicated, however, that other factors may undermine compliance with
contact tracing. Perceived lack of data security and privacy, together
with lack of trust in government, were found to be the main barriers
to adoption \cite{Altmann}. Even in relatively compliant populations,
contact tracing may not be sufficient to control the spread of the
virus. For example, the study \cite{Smieszek} considered the
effectiveness of two different methods of contact tracing within a
closed, and generally compliant population, namely the participants at
a scientific conference on epidemic modelling. One approach was based
on reported contacts, as recorded in a log book, the other based on
the use of unobtrusive wearable proximity sensors. While both methods
were highly tolerated, it was found that neither on its own was able
to give a full picture of meaningful contacts that might have caused
an infection to spread. It is clear that other methods are needed to
control the spread of the virus in addition to contact tracing.

The question addressed in this paper is more modest. We imagine a risk assessment is to be made within a specific workplace on whether and how the workplace can be made `safe' to reopen following lockdown. Here we use the term {\em safe} to refer to the public health of the whole of society, to avoid the workplace contributing to a resurgence of the virus in the general population. Nevertheless, the employees, who are the agents in our model, also benefit from safety, but it is assumed that the individual mortality and morbidity rates are sufficiently low that the payoff to the individual is small. In addition, the 
stakeholders benefiting from the output of workplace, would want
the workplace to be productive. Such stakeholders include general actors who benefit from the upturn in economy, the owners or shareholders of the business in question, and the workers themselves in terms of security of employment. 

Thus, it might seem that
safety and productivity are potentially conflicting aims.
Given a small overall rate of viral infection, an employer might seek to maximise the workplace productivity by staying open, without isolation of exposed workers. But such actions would clearly compromise safety, thus providing as a disbenefit to society. This might be couched in terms of the classical prisoner's dilemma problem within game theory. That is, if every workplace took this attitude, then clearly resurgence within the general population would be probable, whereas one or two isolated `bad apple' 
employers might be able to benefit by maximising
their personal productivity, provided that others don't.  

In fact this `bad apple' principle has been analysed in the context of epidemics by Enright and Kao \cite{Enright}, who ran an agent based simulation where there are precisely such conflicting payoffs. The specific motivation for their study was disease among farm animals where the disbenefit to the farmer of
complying with safety might be particularly harsh, such as the slaughter of her entire herd. They found that a sharp phase transition occurs from sub-critical
$R$ values to super-critical, for a relatively small amount of compliance. Their findings are echoed by those of Eksin {\em et al} \cite{Eksin} in a stochastic network simulation of epidemics. The latter also 
conclude that `a little empathy goes a long way', meaning that 
a focus on treating and isolating those infected can be more advantageous than an approach that seeks to
protect the healthy. See also the recent review \cite{Chang} on the state of the art for game-theoretic approaches to analysing agent behaviour within epidemics. 

The present study was motivated by discussions at the end of April 2020  during a virtual study group workshop on Mathematical Principles for Unlocking the Workplace  at the end of April 2020, organised by the VKEMS\footnote{\url{www.vkemsuk.org}} initiative between the Isaac Newton Institute (Campbridge), the International Centre for Mathematical Sciences (Edinburgh) and the UK's Knowledge Transfer Network.
The participants were split into four teams to discuss different natures of the person-to-person interactions that happen in different working environments; brief verses lengthy, and with a closed set of colleagues versus with an open set of clients. This study arose from  the group look at working environments where interactions were lengthy but with a closed set of colleagues.   

The kind of workplaces we have in mind then, are
those where workers have a relatively static interaction network with other colleagues. Any interactions with other employees or clients can be assumed to socially distanced. An example of such a working environment might be a back
office, which is split into teams that are physically co-located with a number of middle managers and service personnel who naturally migrate between several teams. Another example might be a factory where individual job functions are well demarcated and a typical worker would only need to interact with a relatively small subset of other workers as part of their normal duties. Thus we shall make the simplifying assumption, in what follows, that the workplace can be represented by a fixed interaction network, with each worker as a node, connected by links that 
represent workers who by the nature of their job function cannot effectively socially distance from each other. 

A recently published study \cite{Cuevas} performs agent-based simulations of the spread of the COVID-19 epidemic within a closed `facility' rather than an open population. We note though that that work does not address the specific question addressed here, namely the effect of having some proportion of the population self-isolate, that is remove themselves from the facility, in the case that they have been in contact with someone who develops symptoms. Also, we don't consider a completely closed facility, each of our workers are assumed to go about their daily business outside of the workplace with some base level of infection rate each day. 

In the present work, we imagine a world in which rapid testing, with near instantaneous results and routine contact tracing, is readily available for the whole population. We suppose that there is nonetheless a risk of infection outside of the workplace. We wish to explore the question of the effect within our chosen workplace of measures designed to stop any incoming infection spreading to the entire workforce. In particular, we shall look at the effect of the presence of a proportion of opaque workers, who are not transparent about their infection-risk status, and do not go home when made aware that one of their colleagues has the virus. Is there an incentive for the employers and employees alike to engage fully in transparency? That is, is it valuable to engage with test, trace and isolate procedures in order to halt virus spread while also maximising productivity?  

The rest of the paper is outlined as follows. The subsequent section introduces
our mathematical model, its underlying assumptions and the various parameters that may be tuned to simulate different scenarios. Section \ref{sec:3} contains simulation results under the two scenarios where the underlying rate of infection in the general population is either negligible or significant. We also conduct some approximate mathematical analysis to help explain the results. Section \ref{sec:4} contains discussion of the findings both from a scientific perspective and from the point of view of public policy interventions and workplace psychology. Finally, section \ref{sec:5} makes recommendations. 

\section{The mathematical model}
\label{sec:2}
We have developed a simple discrete-time simulation model posed on a graph representation of an office environment.  An outline of the model is given
in Fig.~\ref{fig:modeloutline}.

\subsection{Underlying assumptions}

\begin{figure}
\centering
\includegraphics[width=0.9\textwidth]{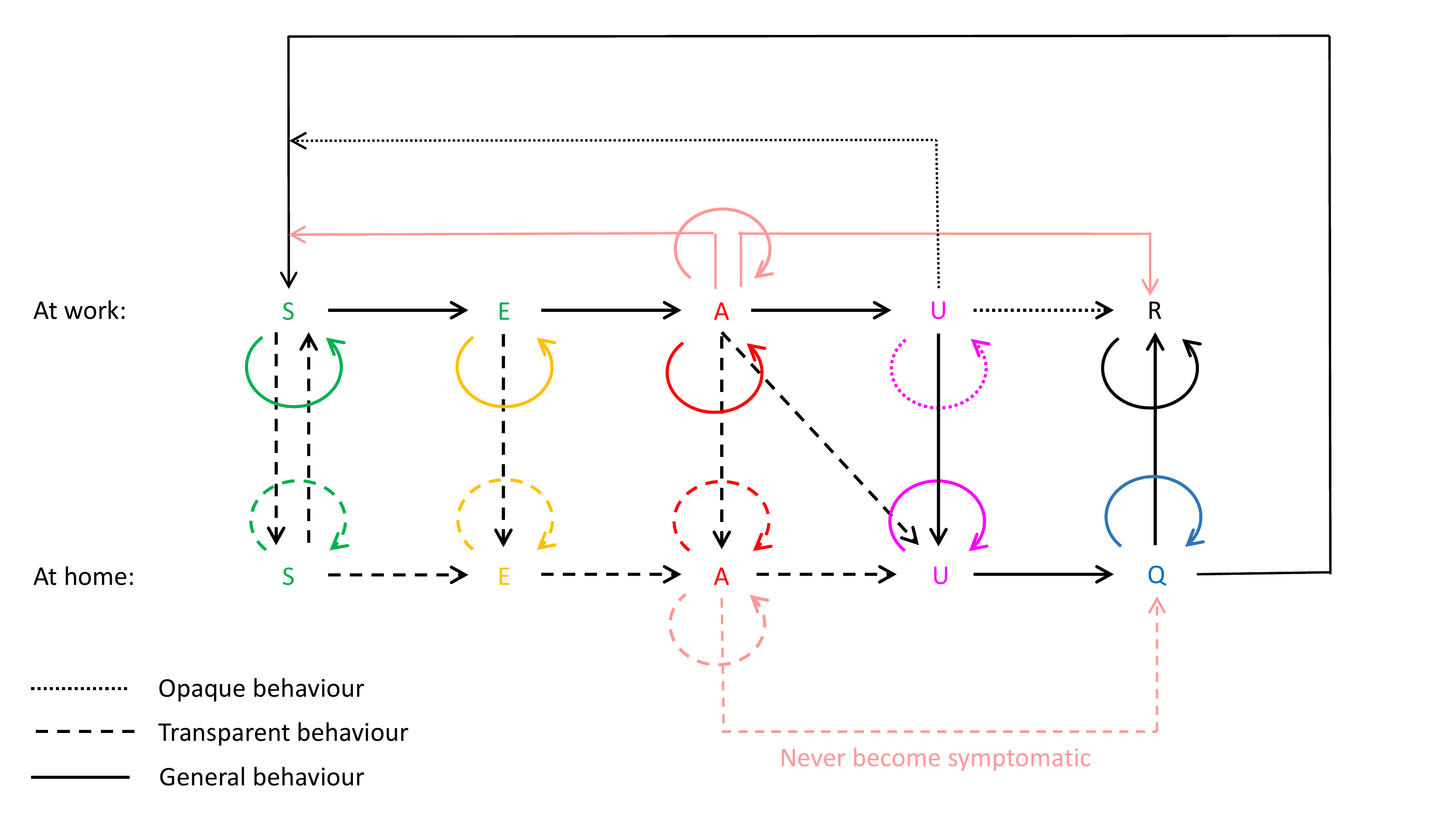}\caption{Sketch outlining the operation of the discrete-event model.  Here $S$ is susceptible, $E$ is exposed (not yet infectious), $A$ is asymptomatic infectious, $U$ is unwell infectious, $Q$ is quarantined and $R$ is recovered (and immune).  Lines/arrows give all possible transitions between states. See text for details.}
\label{fig:modeloutline}
\end{figure}

The basic disease model we choose is a form of 
SEIR model. That is, the usual extension to the Kermack-McKendrick model that allows for four states; ($S$)usceptable, ($E$)xposed (infected but not yet infectious), ($I$)nfectious and ($R$)ecovered (with immunity from re-infection). In line with what is known about COVID-19, in fact we choose a modified variant in which there are two infectious states; $A$ for (A)symptomatic infectious individuals and $U$ for infectious symptomatic, or (U)nwell. We suppose that susceptible individuals can become exposed in one of two ways, either through a (small) 
probability of exposure to the virus present in the general population, or with a much larger probability if one or more of their connected co-workers is infectious. For simplicity, we shall suppose that the rate of exposure outside of the workspace is constant, irrespective of each worker's personal circumstances. For the purposes of the simulation, it will also be
useful to consider a sixth disease state, 
($Q$)uarantined, to represent those who are obliged to be in quarantine post infection. Note that the $Q$ state is only a small subset of those who are isolating at home (see Fig.~\ref{fig:modeloutline}),
which also includes other disease states. 

Thus we can represent the disease dynamics as 
$$
S \to E \to A \to U \to Q \to R,
$$
with the possibility for remaining in each state for periods of time and the possibility of
skipping certain stages altogether (for example, $A$ may transition straight to $Q$ or to $R$ without going through $U$). The full set of possible transitions are illustrated by the arrows in 
Fig.~\ref{fig:modeloutline}.
The specific disease parameters and logical rules that determine the transitions between these disease states are presented in Sec.~\ref{subsec:increment}.

\begin{table}
\centering
\begin{tabular}{|c|c|c|}
\hline 
{\bf Parameter } & {\bf Meaning} & {\bf Value} \\
\hline 
$N$      & number of workers  &    100 \\
$T$      & period of simulation (days) & 100 \\ 
$d$      & mean number of workplace contacts & 3--12 \\
$O$ & percentage of nodes that are opaque & 0--100 \\
\hline
$\alpha$ & probability of community infection  & $\{0, 0.001 \}$ \\
$\beta$ & probability of infection from an infectious connection & 0.1 \\
$\gamma$ & probability of becoming symptomatic  &  0.95\\
$\delta$ & probability of gaining immunity after infection & 0.5 \\
$\epsilon$ & probability of non-transparent isolating due to contact & 0.01 \\
$\zeta$  & probability of non-transparent isolating due to symptoms & 0.01 \\
\hline
$t_E$ & incubation period before infectious & 4 \\
$t_A$ & initial asymptomatic period while infectious & 3 \\
$t_U$ & time following $t_A$ until disease free & 7\\
$t_Q$ & required time of quarantine after symptoms stop & 5 \\
\hline
$\mu_1$ & productivity of home working & $\{0,0.7,1\}$ \\
$\mu_2$ & productivity at work while sick & $\{1,0.2,1\}$ \\
\hline
\end{tabular}
\caption{Parameter values used in the simulations. See text for interpretation and justification.} 
\label{tab:pars}
\end{table}

For convenience, we choose a discrete-time version of the model, in which the fundamental unit of time is the working day. To correct the model for the effect of weekends or other regular workplace closure days, we could in principle exclude such days from our simulation and choose $\alpha$ to be a given function of time, which would be larger after a closure day, and adjust the time intervals $t_{E,A,U,Q}$ accordingly. 

The fundamental model is a dynamic network with $N$ nodes, in which nodes are workers and the state at node $i$ is a 3-tuple: 
\begin{align*}
X_i & =(x_i,p_i,o_i); \\
    & x_i\in \{S,E,A,U,Q,R\},\\
    & p_i \in \{0,1\},\\ 
    & o_i \in \{0,1\} .
\end{align*}
Hence, $x_i$ gives the disease state; $p_i$ is a binary variable that measures whether the worker is present in the workplace ($p_i=1$) or is self-isolating at home ($p_i=0$); 
$o_i$ is a binary variable that determines the {\em opacity} of the worker, namely whether they consistently engage in test, trace and isolate and share their data openly ($o_i=0$), or not ($o_i=1$).  
The workplace contact network of interactions is given by an adjacency matrix
\begin{align*}
A= \{ a_{ij} \}; \quad  & a_{ij} \in \{0,1\}, \\
& \mbox{such that } i,j \mbox { are in contact iff } 
p_ip_ja_{ij}=1.
\end{align*}
Note therefore that while we assume $A_{ij}$ is fixed in time, the actual workplace contact network varies according to whether workers are at home or not. It is useful therefore to define a {\em current workplace contact matrix} $C$ and the set of {\em workplace contacts $W_i$} for each node $i$: 
$$
C=\{c_{ij}\} = \{p_i p_j a_{ij} \} \quad \mathrm{and} \quad 
W_i=\{ j: C_{ij} =1\}.  
$$ 
It is also helpful to define indicator functions
$f_i$ and $g_i$ to 
determine respectively whether one of node $i$'s contacts is infectious, or whether one 
is reporting symptoms:
$$
f_i= \left \{ 
\begin{array}{ll}
1 & \mathrm{if} \: \exists j\in W_i\: \mbox{such that } x_j \in \{A, U\}, \\
0 & \mbox{otherwise} ,
\end{array}
\right .
$$
$$
g_i = \left \{ 
\begin{array}{ll}
1 & \mathrm{if} \: \exists j\in W_i\: \mbox{such that } \: o_j=0\: \mbox{and} \: x_j  = U, \\
0 & \mbox{otherwise} .
\end{array}
\right .
$$
The opacity variable $o_i$ determines an individual worker's behaviour if they become symptomatic or if one of their contacts tests positive for the disease.  

\begin{figure}
\begin{center}
\includegraphics[width=0.32\textwidth]{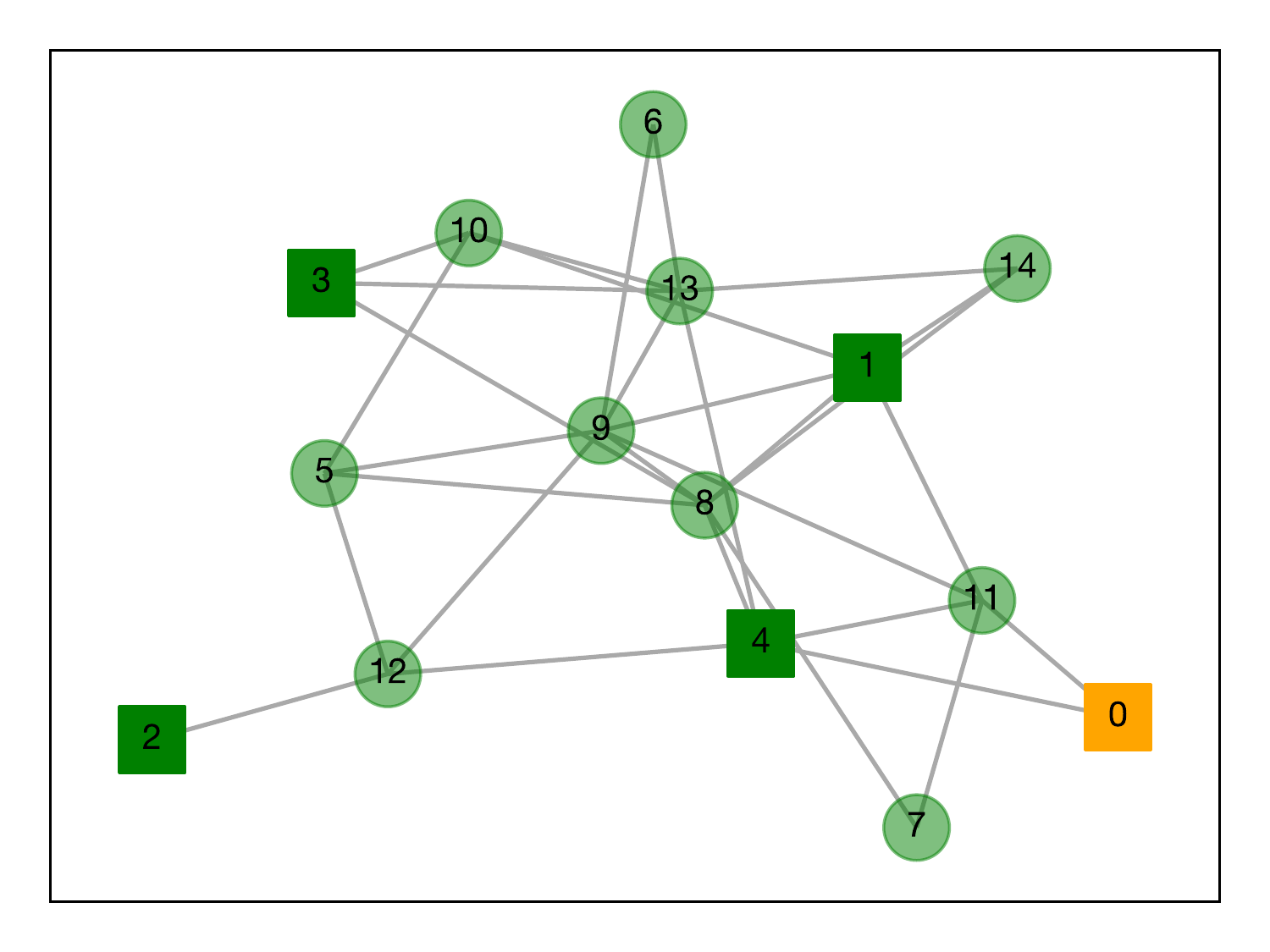}
\includegraphics[width=0.32\textwidth]{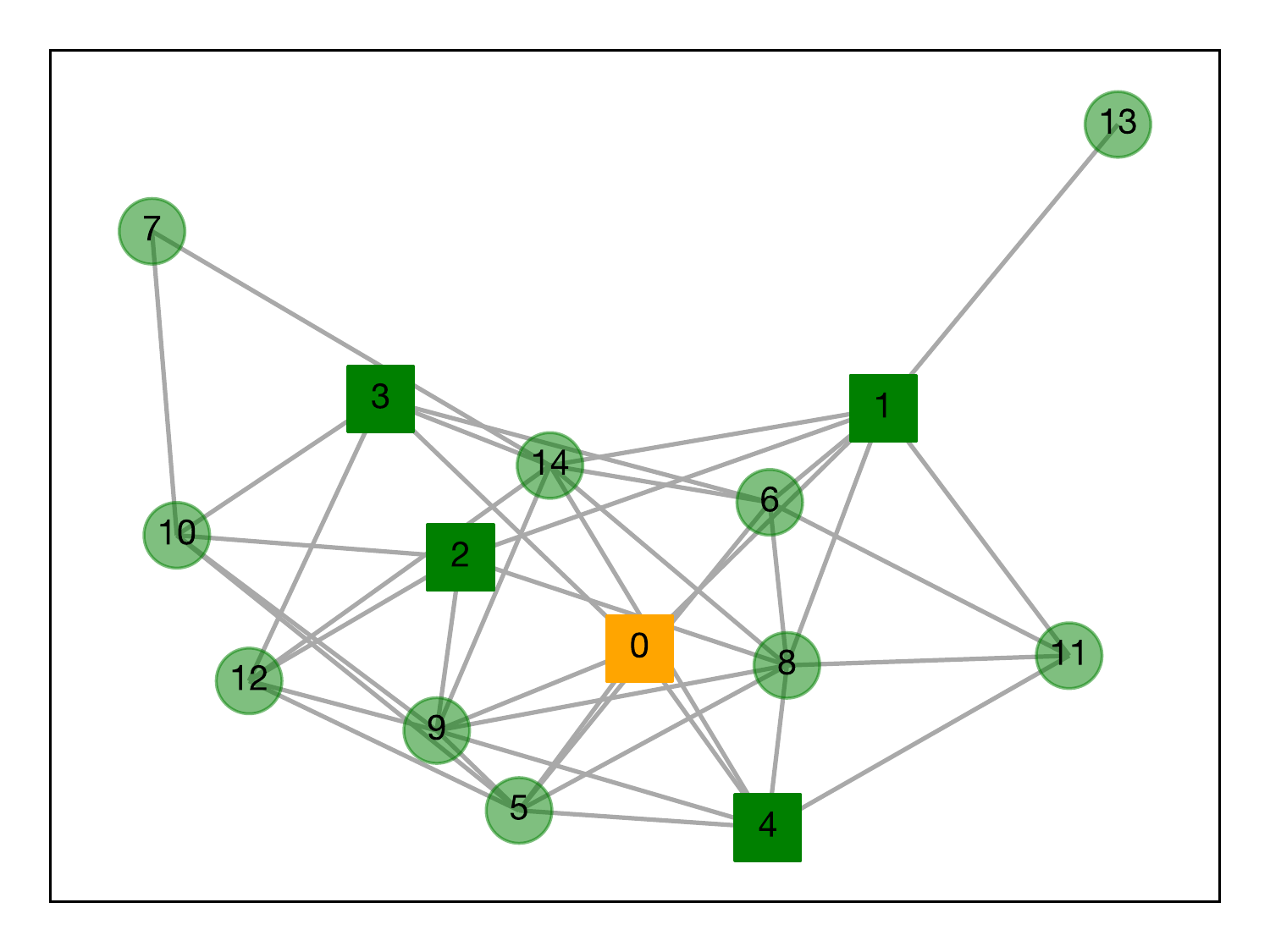}
\includegraphics[width=0.32\textwidth]{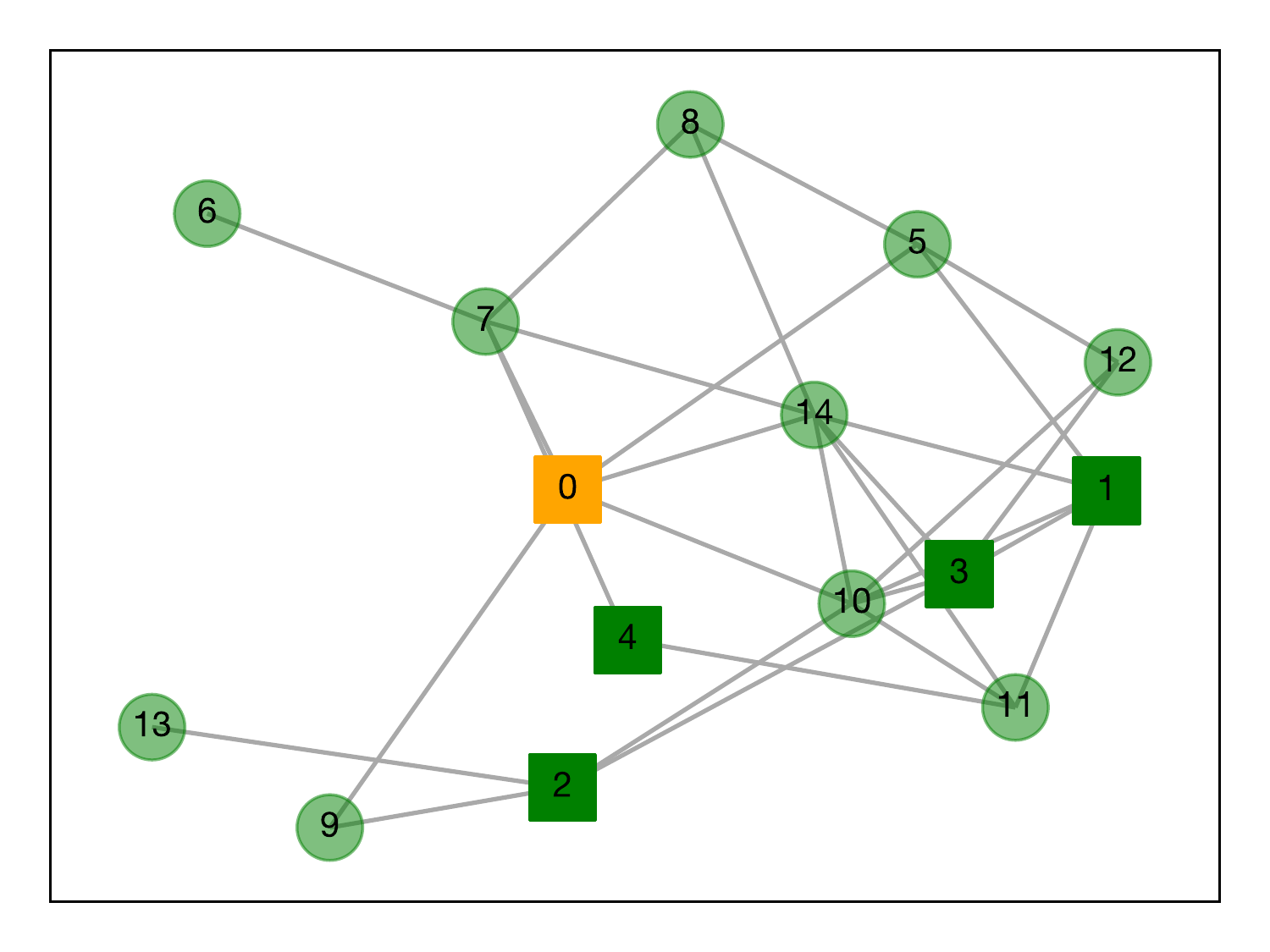}
\includegraphics[width=0.32\textwidth]{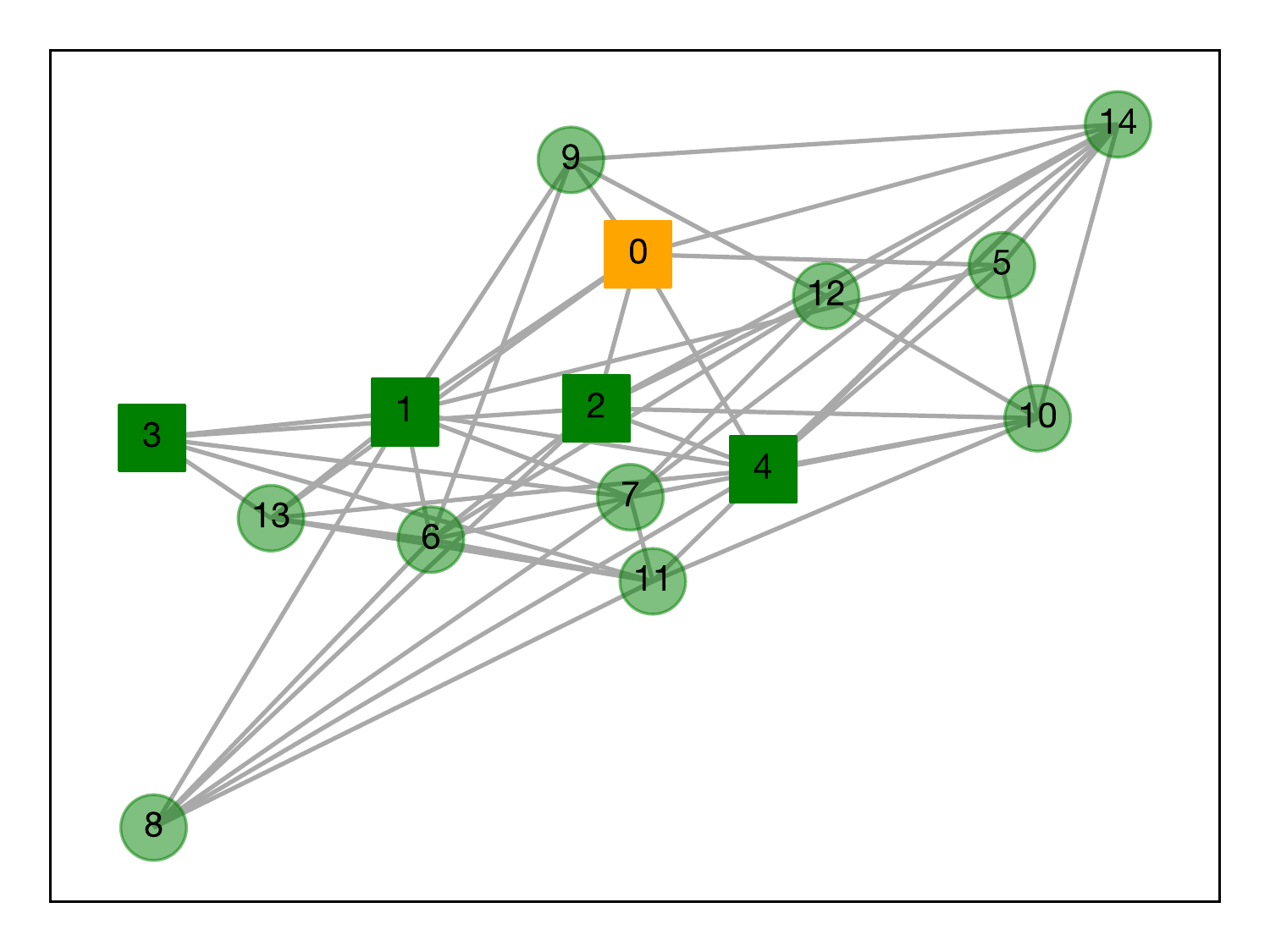}
\includegraphics[width=0.32\textwidth]{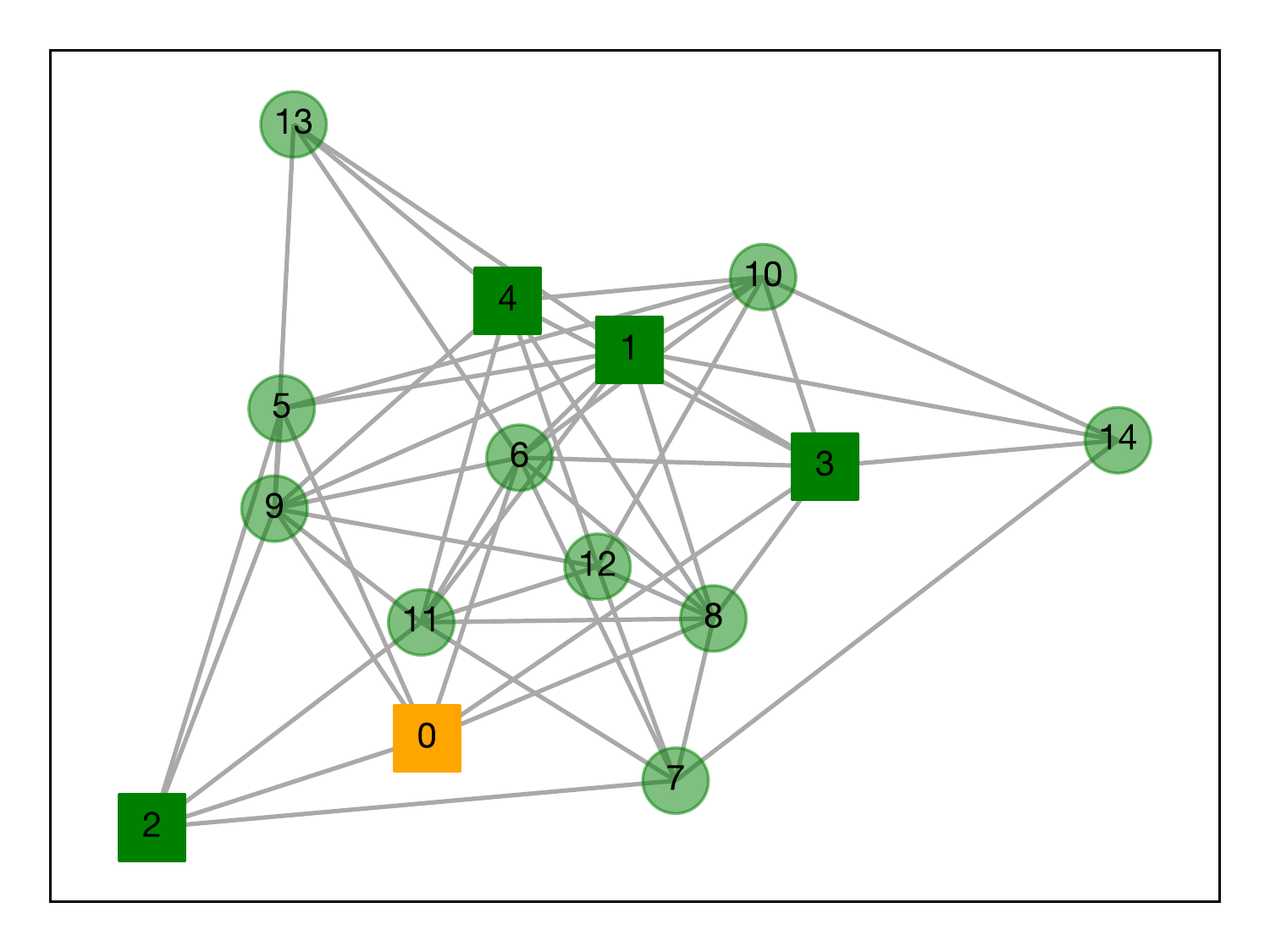}
\includegraphics[width=0.32\textwidth]{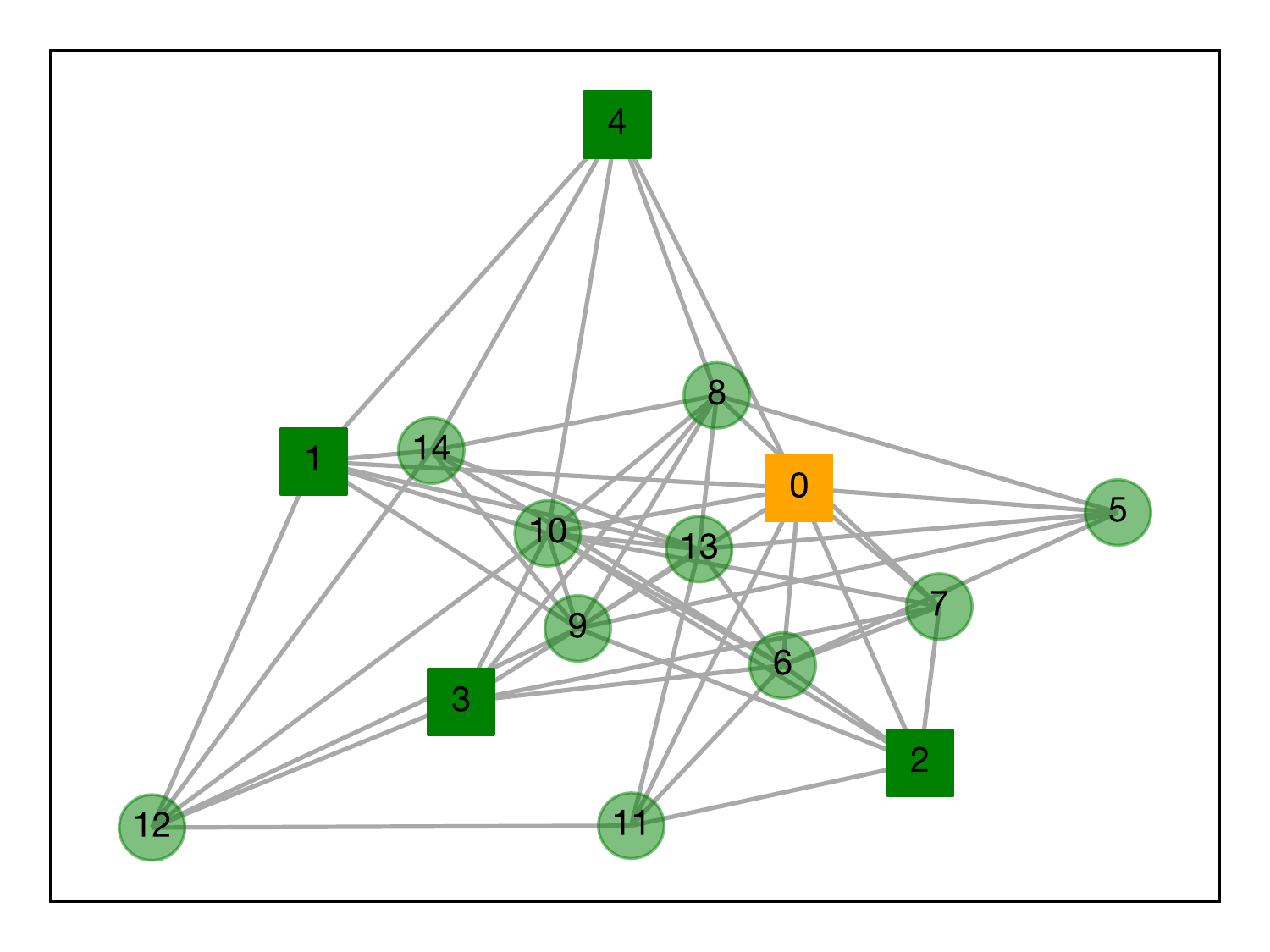}
\caption{Examples of randomly generated workplace contact networks,
between workers for six different random initial conditions of a network of $N=15$ with  having a $1/3$ of the workforce opaque (solid square) and $2/3$ 
transparent (circle). One individual is assumed to be infected with the virus and is in state $E$ (the orange node labeled 0) whereas all others are susceptible (green). 
(Top) realisations with 
$d=4$; (bottom) $d=8$.} 
\label{fig:network}
\end{center}
\end{figure}

\subsection{Time increment}
\label{subsec:increment}

The key to the model is the time update step, which determines how each worker transitions between disease states.  We first delineate all the update rules for transparent workers, before explaining only what is different in the case of an opaque worker. 

\paragraph{The behaviour of transparent workers:} Assume $o_i=0$, and consider the transition taking place in one time step to worker $i$. 
\begin{enumerate}
\item {\bf Infection:} If worker $i$ is in work and susceptible, such that $p_i=1$ and $x_i=S$, then the probability of infection (transition to disease state $E$) in the next time step is $ \alpha + \beta f_i $. Otherwise
$x_i$ remains as $S$. 
\item {\bf Isolation because of contact:} In addition, if worker $i$ is in work, and a transparent work contact is unwell, such that $p_i=1$ and $g_i=1$, then
$p_i \to 0$ in the next timestep. That is, worker $i$ goes home, irrespective of its own disease state $x_i$ in the present time instant or the next. 
\item {\bf Exposure:} If worker $i$ is exposed, such that $x_i=E$, then $x_i$ remains in this state for a total number of contiguous days equal to $t_{E}$, before transitioning to disease state $A$, as $i$ becomes infectious but asymptomatic.
\item {\bf Asymptomatic infectiousness:} If $i$ is asymptomatic, such that $x_i=A$, then $x_i$ remains in this state for a total number of contiguous days $t_A$. After this, $x_i$ transitions, in the next timestep and with probability $\gamma$, to $U$. Otherwise, $x_i$ remains in disease state $A$ for a further $t_U$ days, without ever passing to $U$. In other words, there is a probability it is possible that worker $i$ never feels unwell.
\item {\bf Symptomatic infectiousness} 
If worker $i$ is unwell, such that $x_i=U$, then $x_i$ remains in this state for a total of $t_{U}$ contiguous days, after which worker $i$ passes to the recovery step.
\item {\bf Isolation due to symptoms:} 
In addition, if worker $i$ is in work and unwell, such that $p_i=1$ and $x_i=U$, then $p_i \to 0$ for the current timestep. 
That is, as soon as an individual becomes unwell, they do not come into work (provided that they are transparent). 
\item {\bf Recovery:} After $t_A+t_U$ time steps since the transition to disease state $A$, worker $i$ becomes disease free.
If $p_i=0$, then $x_i$ transitions to $Q$ as $i$ goes into a 
post-symptomatic quarantine state, in which it remains for a total of $t_Q$ consecutive timesteps, after which $i$ goes to the immunity step.
Else, that is if $p_i=1$, worker $i$ passes to the immunity step. 
\item {\bf Immunity:} At the end of the infection, with probability $\delta$, the worker develops
viral antibodies so that $x_i\to R$ and $p_i\to 1$. 
Else, with probability $1-\delta$, the worker does not become immune, so that $x_i\to S$ and $p_i\to 1$.
If $x_i = R$ at any timestep, irrespective of the value of $p_i$, then worker $i$ remains in $R$ for the rest of the simulation, and $p_i$ remains 1. 
\end{enumerate}

\paragraph{The behaviour of opaque workers.}
Workers who are not-transparent, $o_i=1$, are assigned at the beginning of the simulation and remain that way for all time steps. 
Their behaviour is identical to that of transparent workers except
\begin{description}
\item[$\quad 2^\prime$.]  If $p_i=1$ and $g_i=1$, then
$p_i \to 0$ in the next timestep, with probability $\epsilon$.
Otherwise $p_i$ remains 1. 
That is, even if present opaque workers have unwell contacts, they might stay in work.
\item[$\quad 6^\prime$.]
If $p_i=1$ and $x_i=U$, and $x_i=A$ in the previous timestep, then $p_i \to 0$ for the current timestep  with probability $\zeta$. 
Otherwise $p_i$ remains 1 for the entire time $t_U$.
So when opaque workers become unwell, they may not go home and thus their contacts are unaware that they may be infected and also do not go home.
\end{description}

\section{Simulation results}
\label{sec:3}

All simulations were carried out in python using the parameter values given in Table \ref{tab:pars}. These values are not supposed to accurately fit to a particular outbreak, but are broadly inspired by COVID-19, so that the length of an infection, following incubation is $t_A+t_U=10$ working days. Note that, due to reported low death rates among otherwise healthy working age populations, we have simplified by assuming that all workers are eventually sufficiently healthy to return to work at the end of the infection; this may be assumed to be a ``best case scenario''. We have further simplified by assuming that incubation and infection times $t_{E,A,U}$ are
deterministic, whereas a more representative simulation would allow these parameters to be chosen from a distribution. We have also supposed that no individual is vaccinated. 

At the time of writing, it is not clear what proportion of individuals
obtain immunity having had the disease, so we make the reasonable
assumption that $\delta=50\%$ of infectious individuals develop
immunity, which lasts for the rest of the simulation run time. The
incubation period before infectiousness, the degree to which
individuals are infectious before they develop symptoms, and the
proportion of individuals that are infectious but never develop
symptoms have also not been clearly established. Thus the relevant
parameter choices $t_E=4$, $t_A=3$ and $\gamma=0.95$ are intended to
be illustrative of what might be the case. Note that the benefit of
transparency is particularly sensitive to the choice of $\gamma$, the
probability of developing symptoms. Although this value is at the
lower end of the current estimate of $4\%-41\%$ of COVID-19 positive
patients being asymptomatic \cite{Byambasuren}, the chosen values of
$\gamma$ can equally be interpreted within the model as there being a
probability of $0.95$ of an infectious patient being detected within
three working days through a combined regime of regularly testing and
reporting of suspicious symptoms.

Nevertheless, the parameter values can readily be altered as more information becomes available for COVID-19, or indeed to model any other national-scale epidemic.

We have chosen a workplace with $N=100$ employees. In each run, the topology of this workplace is generated as a random (Erd\"{o}s-Renyi) symmetric graph with probability $d/N$ that each edge $A_{ij}=1$, independent of other edges. By definition we choose $A_{ii}=0$, hence the average degree of each graph is actually 
$$
(N-1)\frac{d}{N} = 0.99 d.
$$
For a given opacity, $O$, we use the same process of ensuring that node $i$ has probability $O$ that
$o_i=1$ independently of the 
value $o_j$ for any other node $j$.
Figure \ref{fig:network} illustrates examples of graphs that are generated in this manner.

When analysing the results of simulations, it is useful to have a measure of productivity
\begin{equation}
\mbox{Productivity}  = \frac{1}{NT}\sum_{0<t<T}\left [
\mathbb{ 1}_{\{x_i \neq U, p_i=1 \} } (x_i)
+ \mu_1 \mathbb{1}_{\{x_i = U, p_i=1 \} } (x_i)
+\mu_2  \mathbb{1}_{\{x_i \neq U, p_i=0 \} } (x_i)
\right ]
\label{eq:Prod}
\end{equation}
and lost productivity
\begin{equation}
\mbox{Productivity Deficit}  = 1- \mbox{Productivity}. 
\label{eq:Pdef}
\end{equation}

\begin{figure}
\begin{center}
\includegraphics[width=0.3\textwidth]{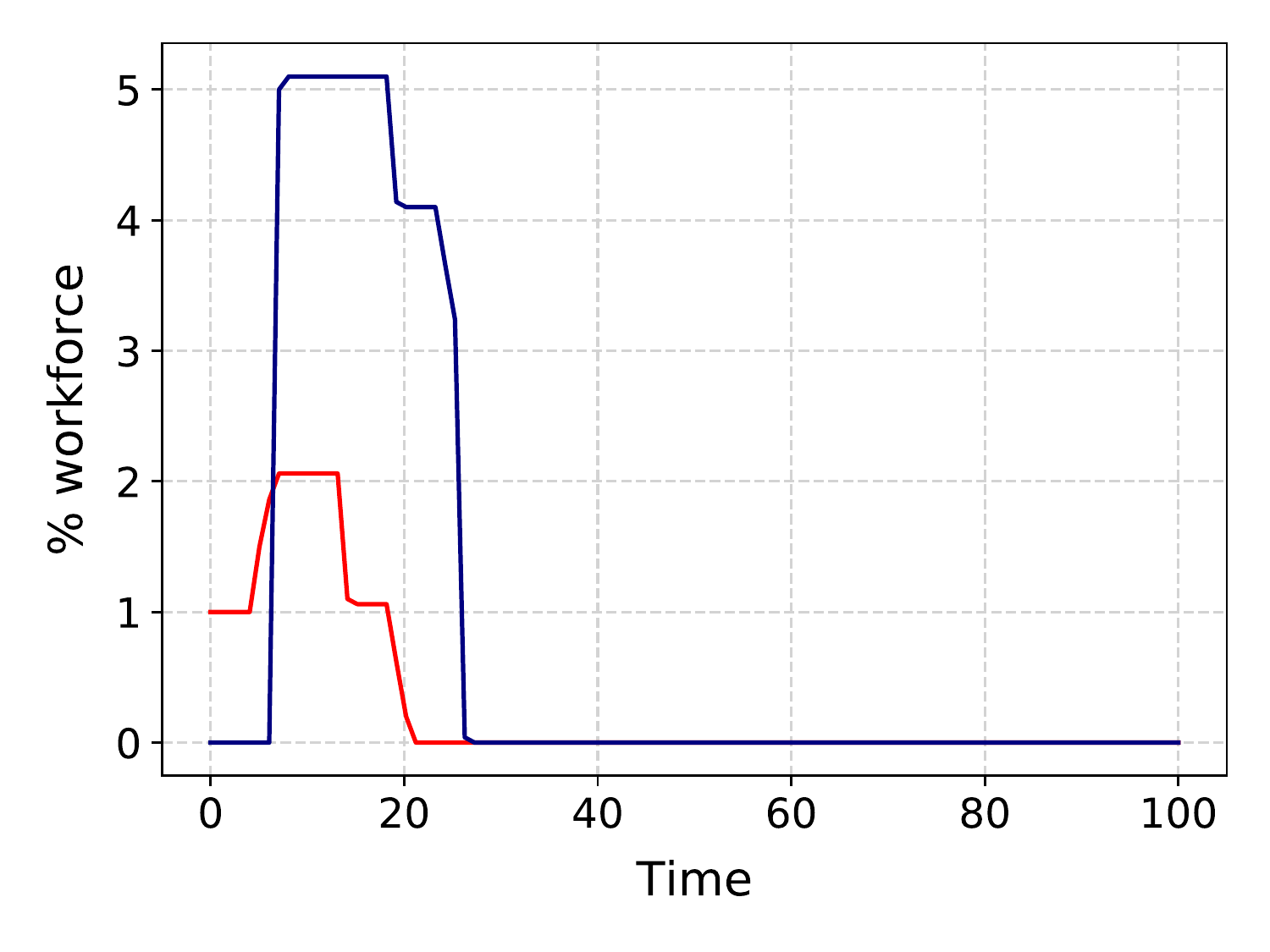}
\includegraphics[width=0.3\textwidth]{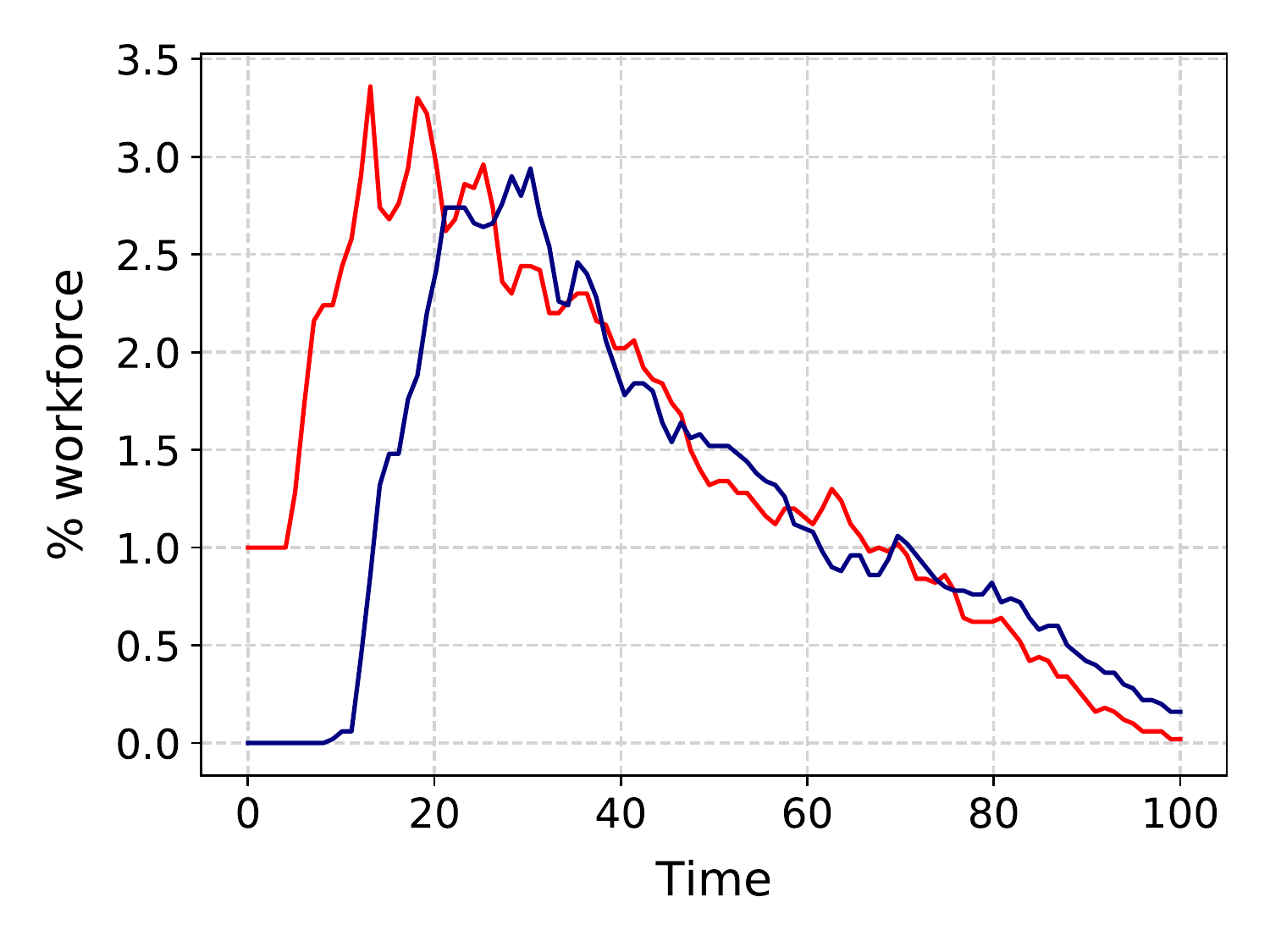}
\includegraphics[width=0.3\textwidth]{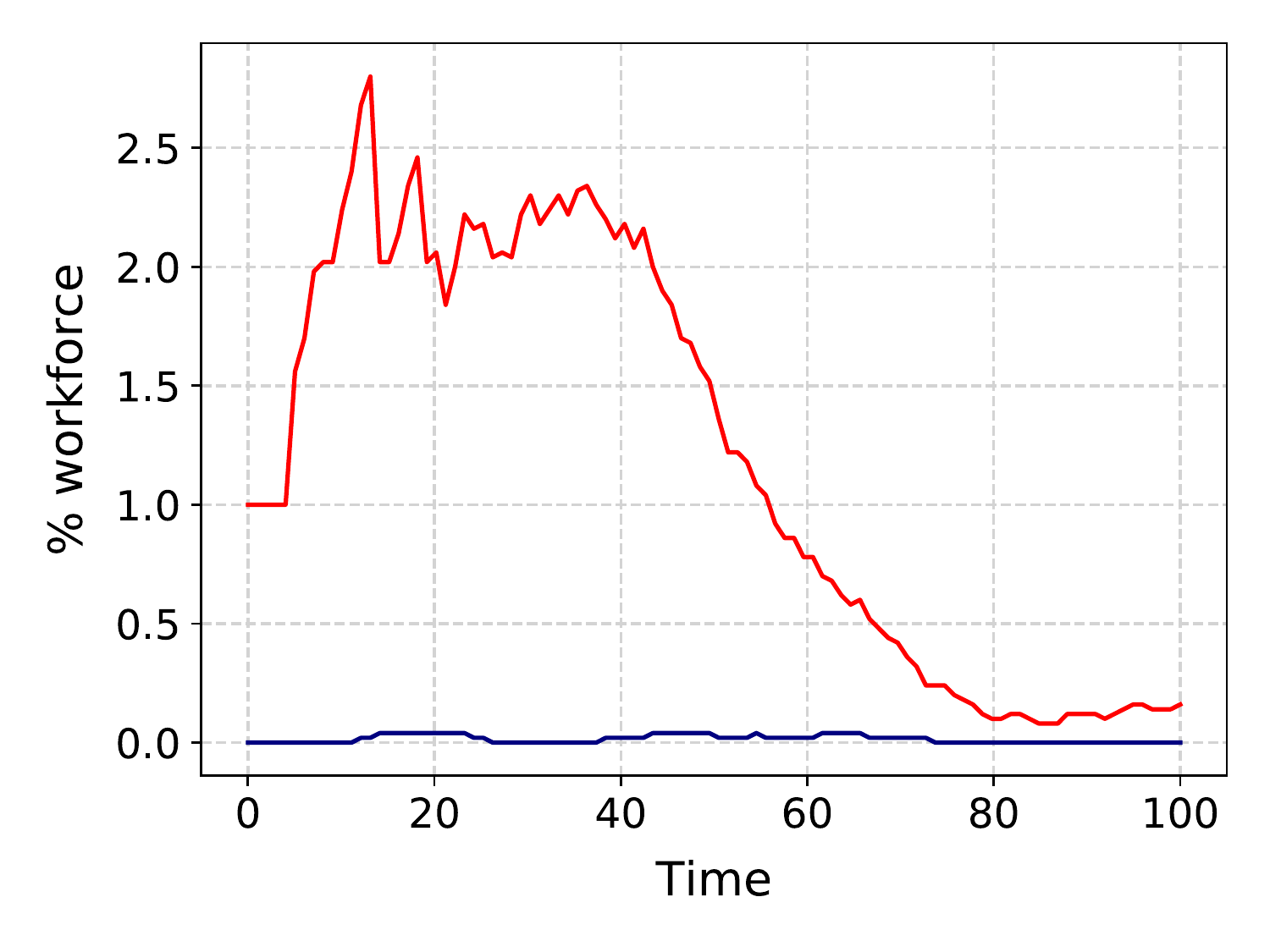}
\includegraphics[width=0.3\textwidth]{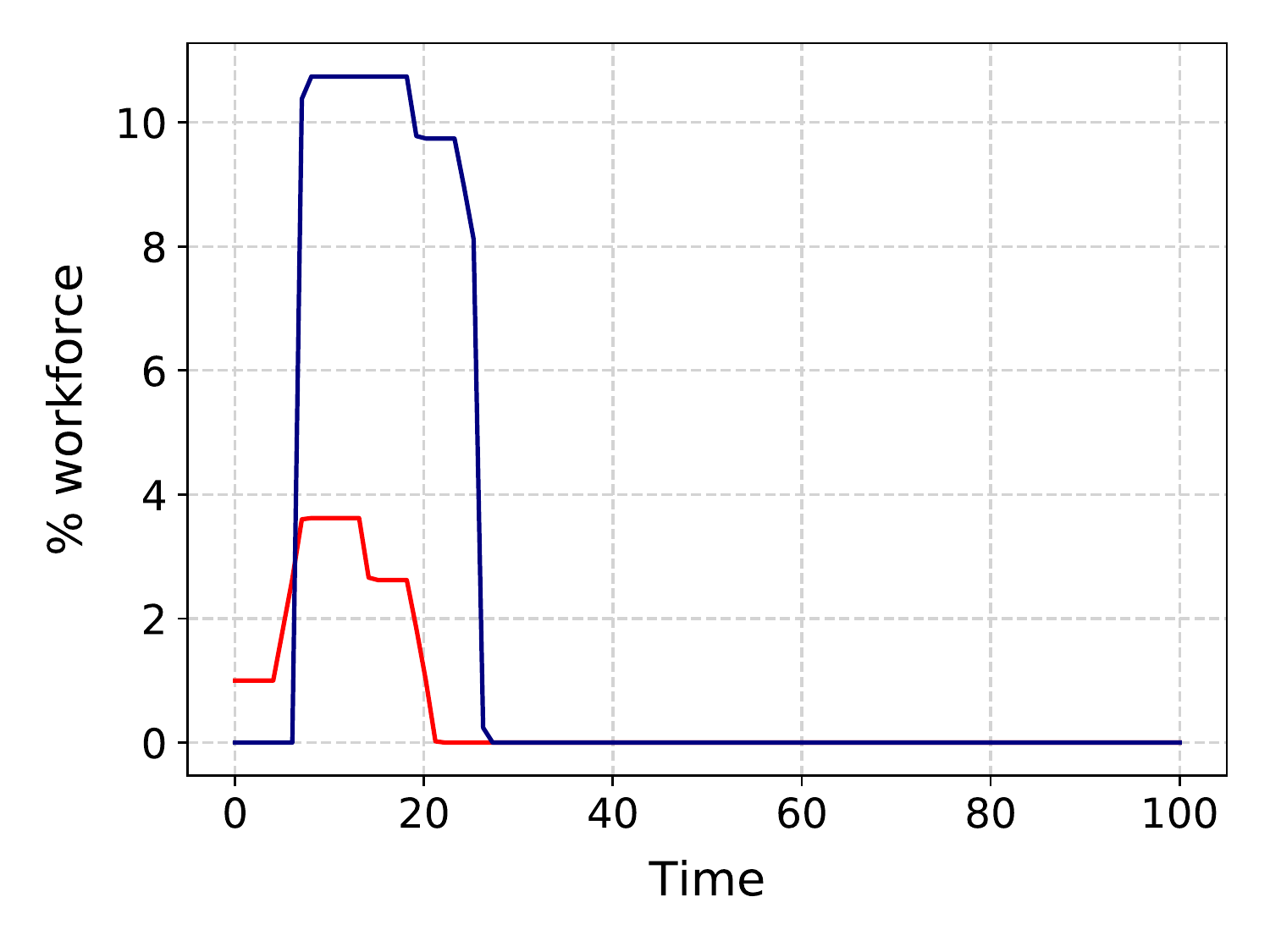}
\includegraphics[width=0.3\textwidth]{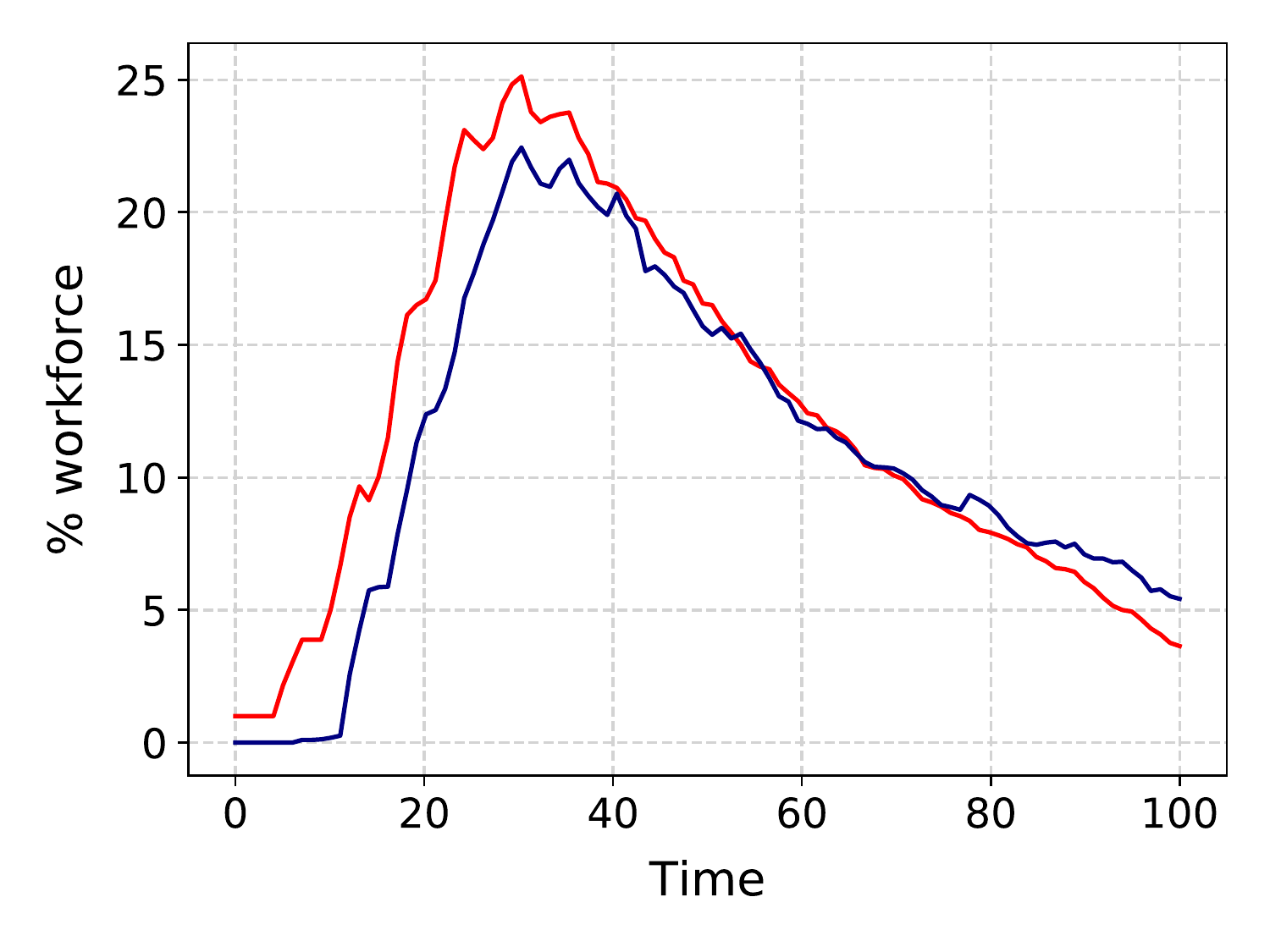}
\includegraphics[width=0.3\textwidth]{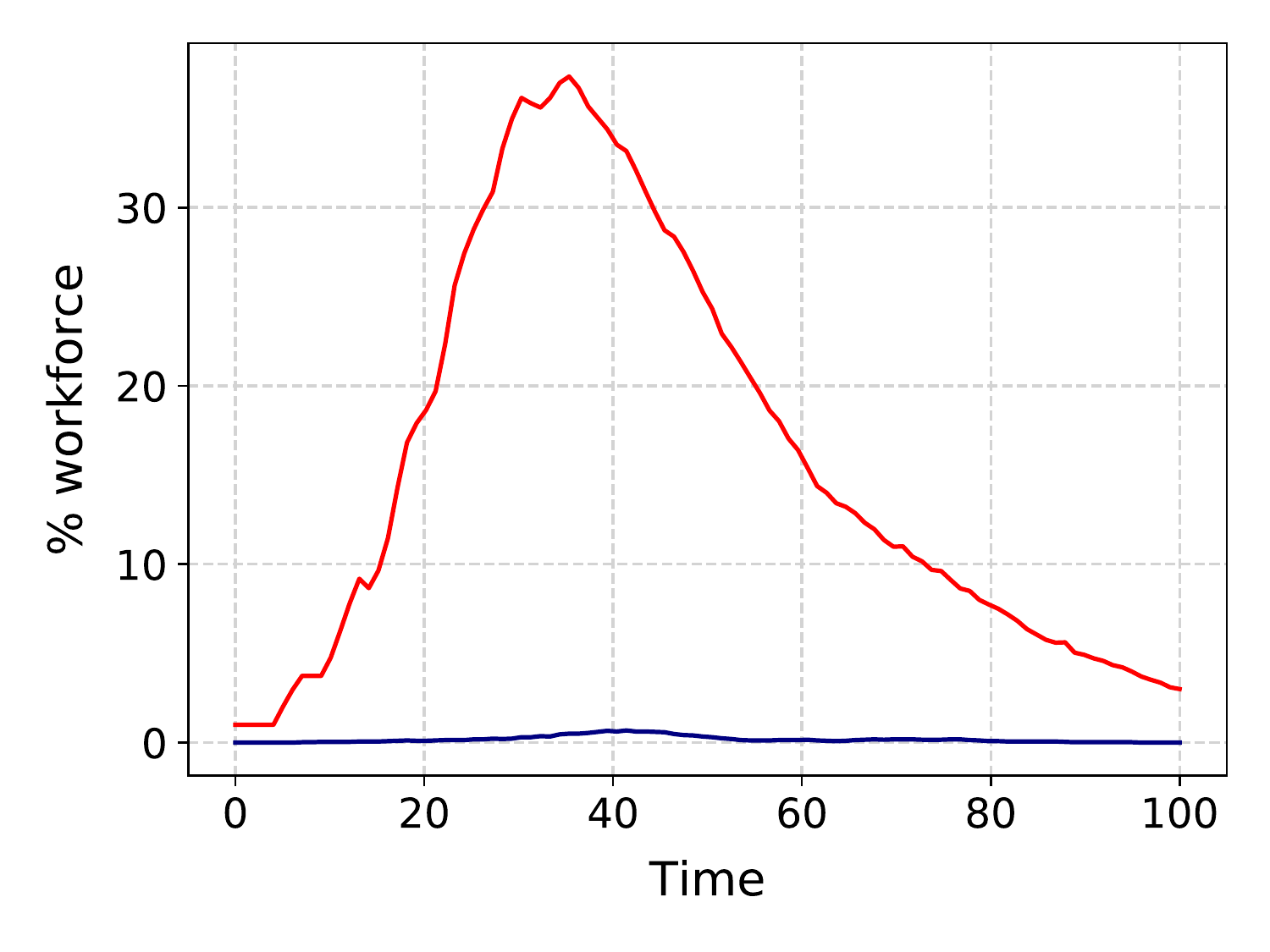}
\caption{Examples of cold runs for different size of contact network $d$ and degree of opacity $O$ against time. 
In each plot, the red curve gives the proportion of unwell individuals (in state $U$) and the blue curve gives the proportion who are at home (with $p_i=0$).  Parameter values are: (top left) $d=4$, $O=0$; (top middle) $d=4$, $O=50\%$;
(top right) $d=4$, $O=100\%$; (bottom left) $d=10$, $O=0$; (bottom middle) $d=10$, $O=50\%$;
(bottom right) $d=10$, $O=100\%$.}
\label{fig:cold1}
\end{center}
\end{figure}

The rationale behind the parameters $\mu_1$ and $\mu_2$ is that those in work are assumed to be fully productive 
if not sick and have fractional productivity $\mu_1$ if sick, whereas those isolating at home 
do no work if they are sick and 
have fractional productivity $\mu_2$ if not.
In what follows, we shall take two extreme and one balanced measure of productivity.
\begin{equation}
\mbox{``academic": } \mu_1=0, \: \mu_2=1; \quad
\mbox{``factory": } \mu_1=1, \: \mu_2=0; 
\quad
\mbox{``office": } \mu_1=0.2, \: \mu_2=0.7. 
\label{eq:workplace}
\end{equation}

When running simulations, we shall consider two cases, which we refer to as {\em running cold} and {\em running hot} depending on whether the overall rates of infection in society are negligible or not. 
\begin{equation}
\mbox{``running cold": } \alpha=0; \qquad \mbox{``running hot": } \alpha=0.001.  
\label{eq:hunt_the_thimble}
\end{equation}

\subsection{Running cold}
In a cold run, the chance of an infection from the outside world is negligible, so that the parameter $\alpha=0$. 

We start simulations at time zero
with one exposed individual and all other individuals in state $S$ (we assume that the state of the disease in the general population is that there is a negligible number of individuals who are already immune). Examples of such simulations 
are shown in Fig.~\ref{fig:cold1}. 

Note from the simulations the benefit of transparency. The left hand-plots show how an infection that starts with one individual at day $0$ quickly dies out. 
Whereas if 
$50\%$ or $100\%$ of the workers are opaque (middle and right panels) the consequence of that initial infection is still present in the workplace 
after $100$ days. Note how the size of the contact network (difference between $d=4$ in upper panels and $d=10$ in lower panels) makes little qualitative difference. However there is a massive quantitative difference when not all workers are transparent. Given 50$\%$ opacity, the maximum size of the outbreak is such that with $d=10$ there are about $25$ people who are sick at around day $30$, rising to almost $40$ people with 100\% opacity. 

\begin{figure}
\centering
\includegraphics[width=0.45\textwidth]{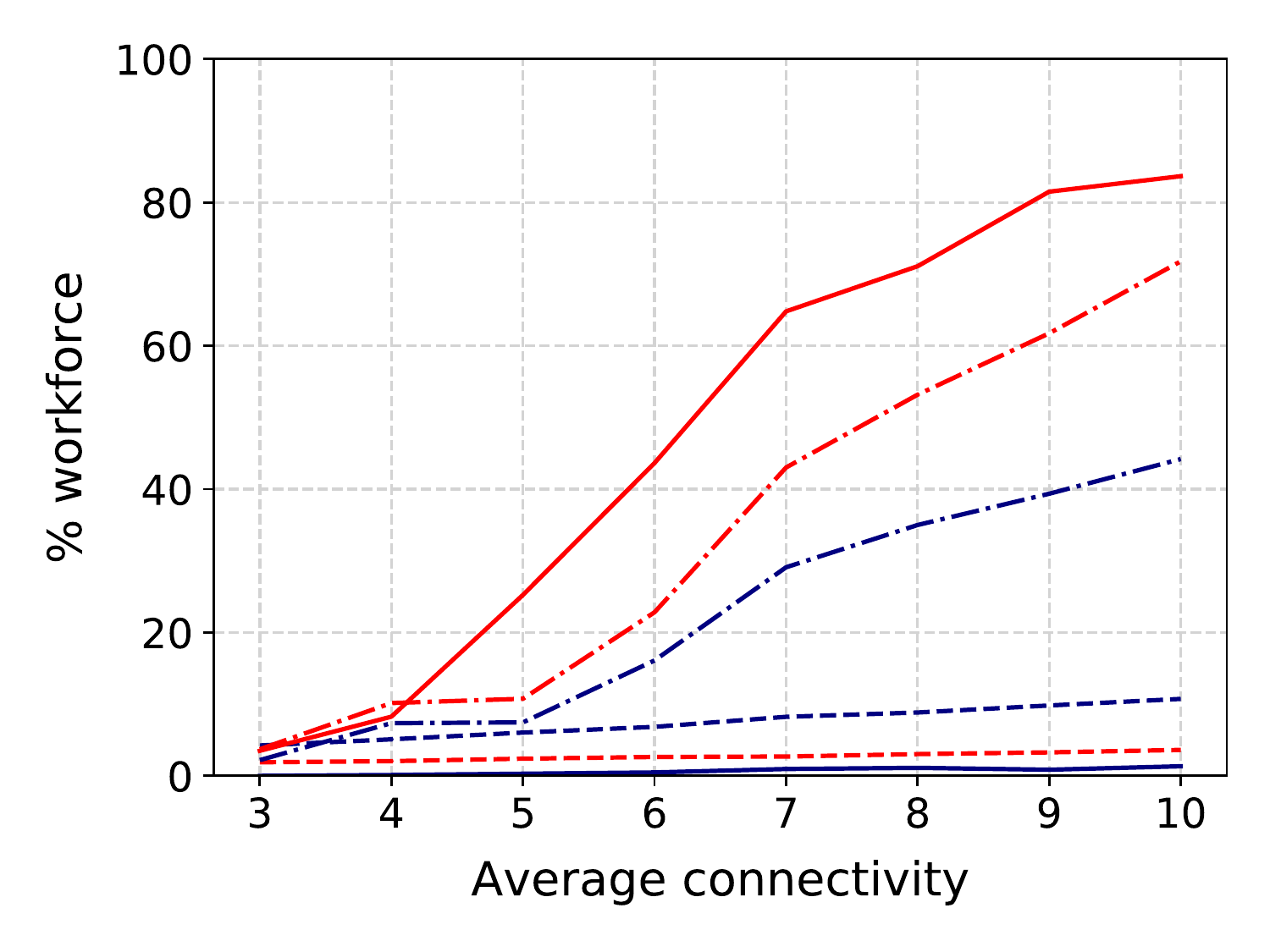}\includegraphics[width=0.45\textwidth]{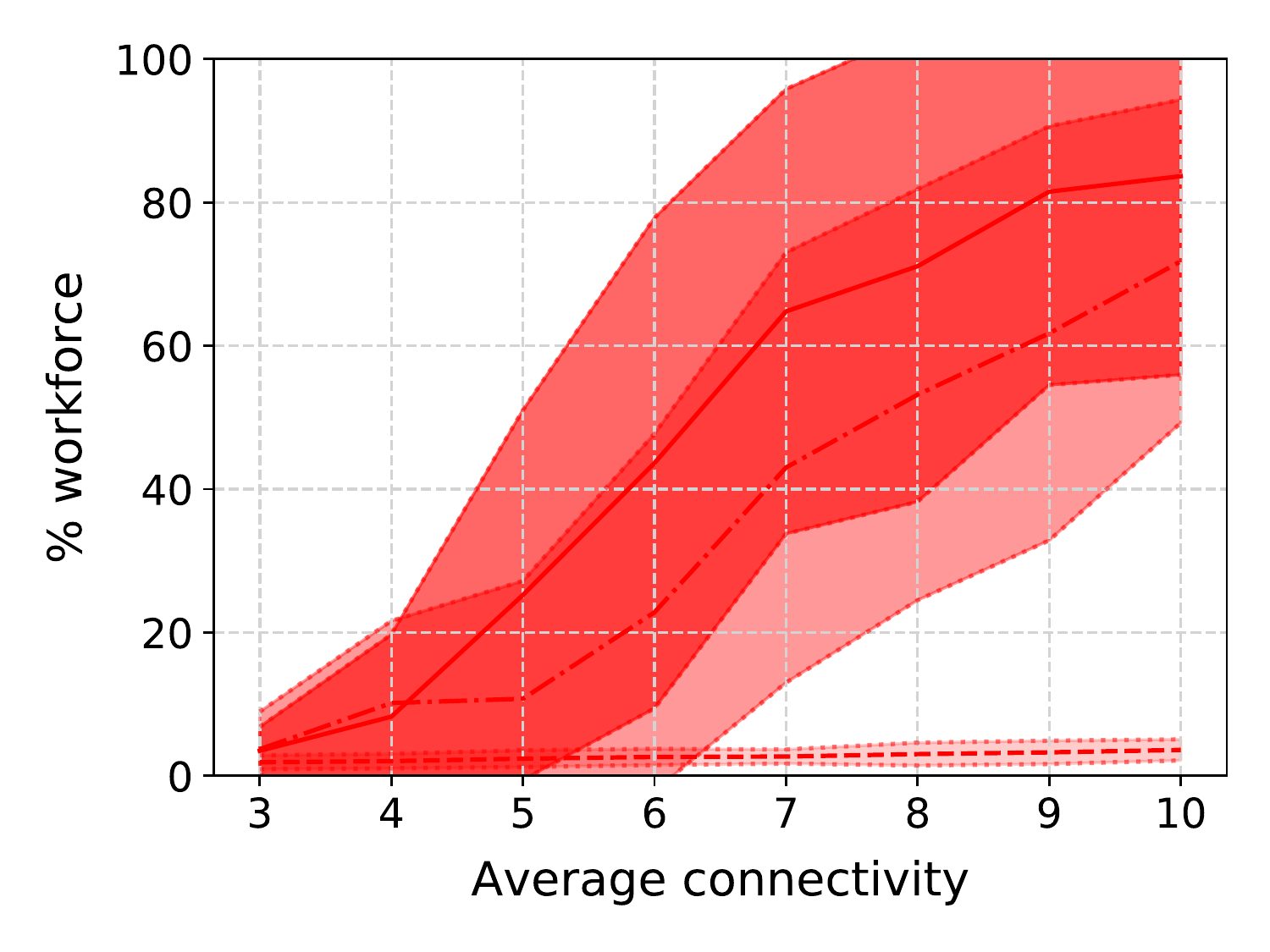}
\caption{(Left) Proportion of workforce that become unwell at least once (red lines) or go home (blue lines) as a function of the average contact network size $d$. Results are shown for three different opacities $O=0\%$ (dashed lines),  $O=50\%$ 
(dot-dashed) and $O=100\%$ (solid). (Right) Mean, plus and minus one standard deviation of the becoming ill proportions.}
\label{fig:cold2}
\end{figure}

\begin{figure}
\centering
\includegraphics[width=0.45\textwidth]{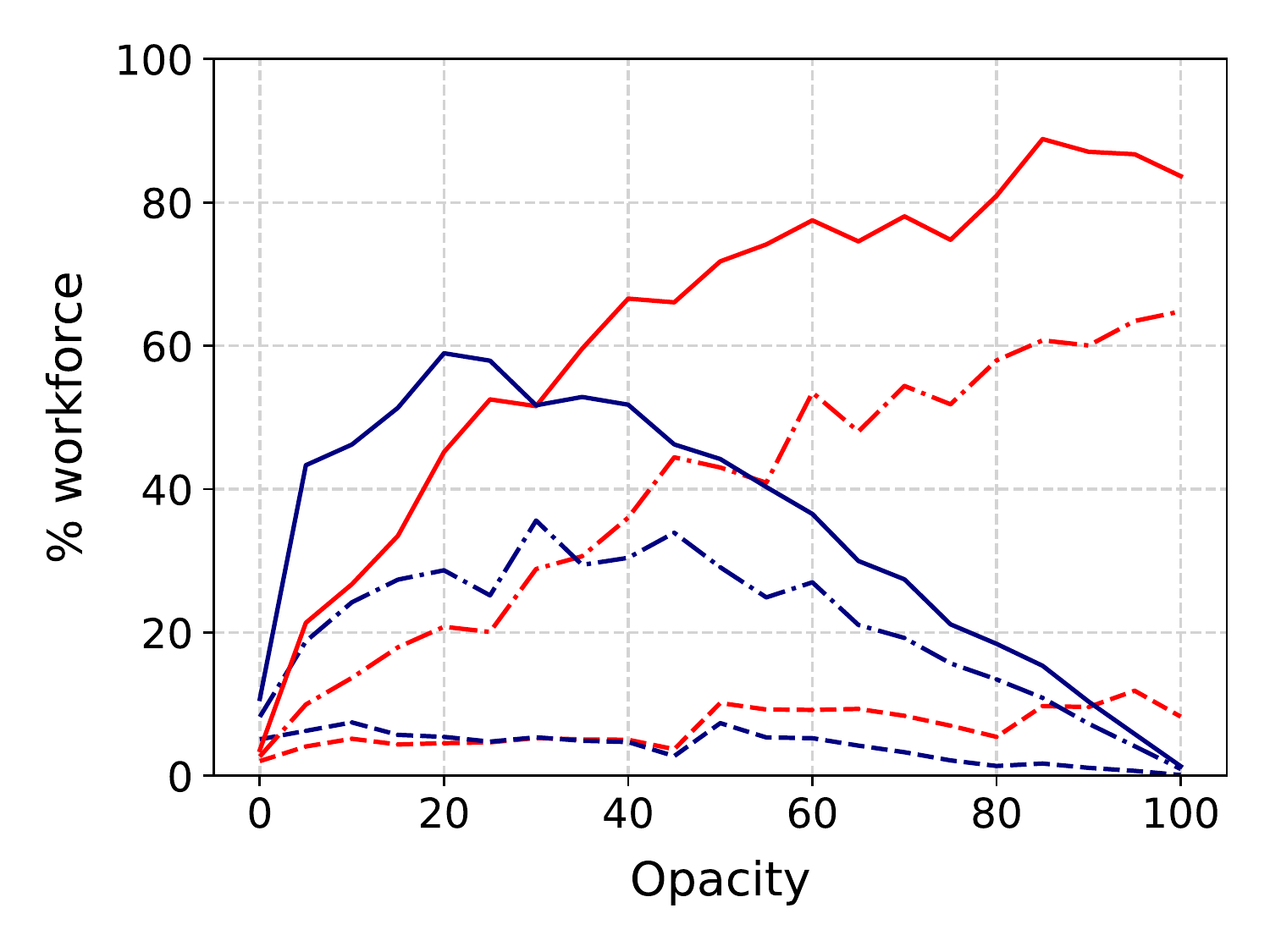}\includegraphics[width=0.45\textwidth]{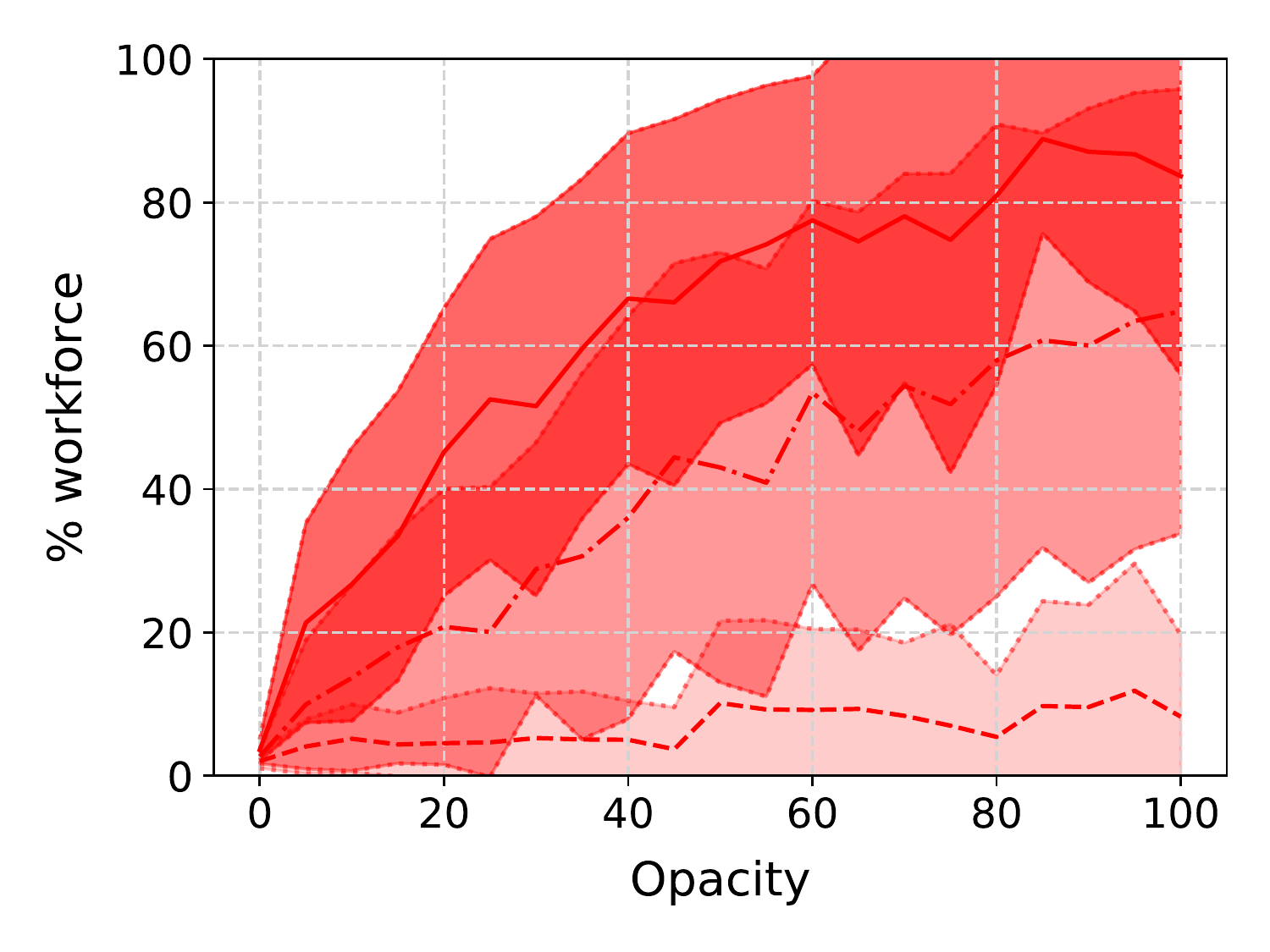}
\caption{Similar to
Fig.~\ref{fig:cold2} but showing variation with opacity for three different values of average degree; $d=4$ (dashed lines), $d=7$ (dot-dashed) and $d=10$ (solid).}
\label{fig:cold3}
\end{figure}

In addition to single runs, it is useful to generate a statistical ensemble. We have done this for a range of $d$ and $O$ values, taking 50 repeats for each parameter value. The results are shown in Figs.~\ref{fig:cold2} and \ref{fig:cold3}, which show the proportion of workers that become unwell (enter state $U$) and the proportion that go home (have $p_i=1$) at least once during the 100 day simulation time. 

Observing the results against $d$ (Fig.~\ref{fig:cold3}), we note how, for the $100\%$ opacity case with degree $10$, almost all of the workforce appear to catch the disease. 
Thus a certain amount of herd immunity is established in the population (recall the immunity rate $\delta=0.5$) and this is what causes the infection rate to decrease towards
the end of the simulation. The results for $50\%$ opacity are similar. Smaller contact network sizes however result in smaller infections.  

Further conclusions can be drawn from the graphs plotted against opacity in Fig.~\ref{fig:cold3}. Here note that, for the case of the highest degree, there is a sharp increase in 
the proportion of infected individuals for low opacity. 
For an intermediate degree, the proportion of infected individuals appears to vary more linearly with opacity, 
with the sharp increase, if there is one, occurring later, perhaps around 20\% (although notice the large standard deviation).  For the lowest degree value, the number of infected individuals appears to be low, with the sharpest 
increase at around $45\%$.

\begin{figure}
\begin{center}
\includegraphics[width=0.45\textwidth]{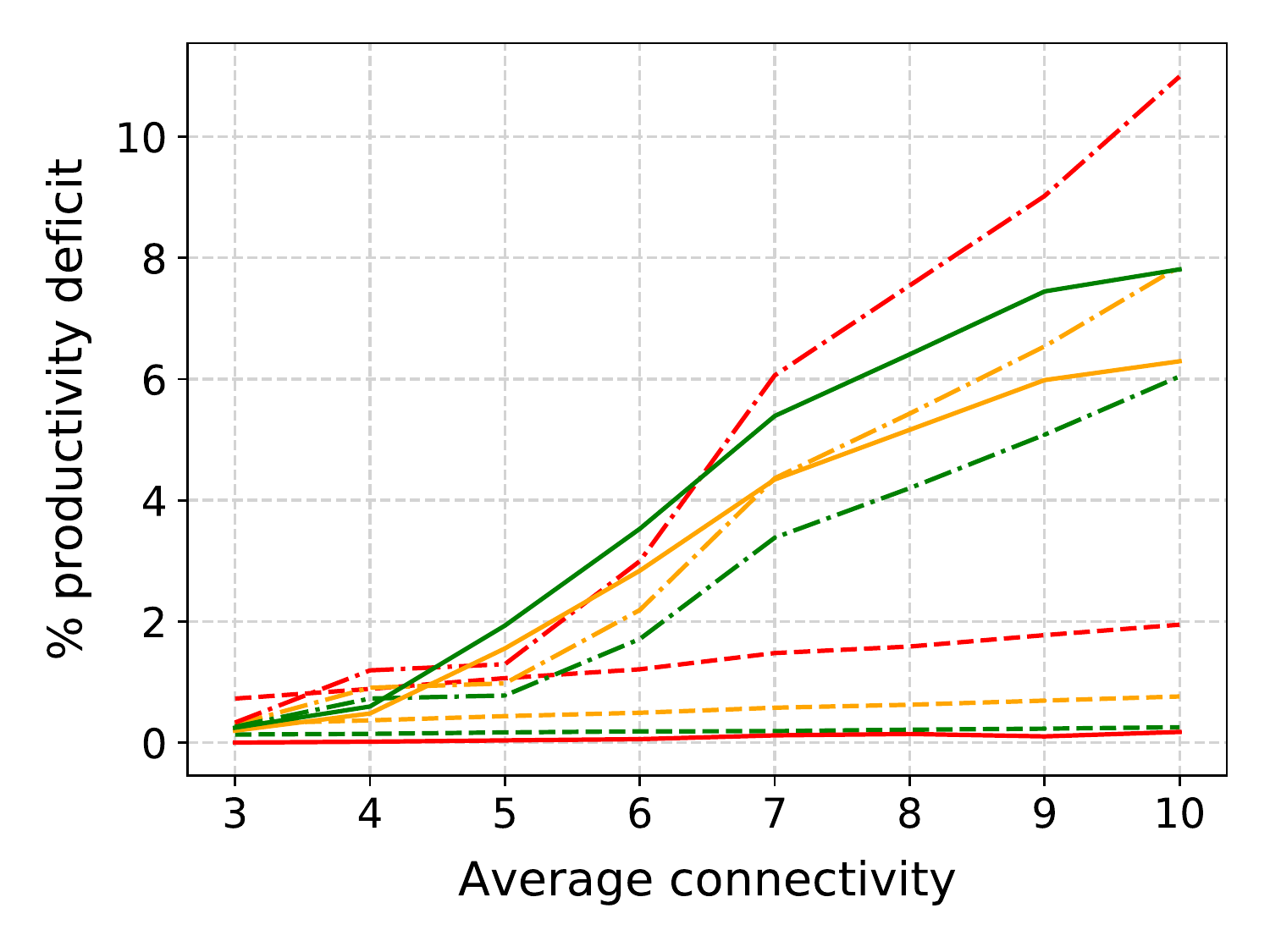}
\includegraphics[width=0.45\textwidth]{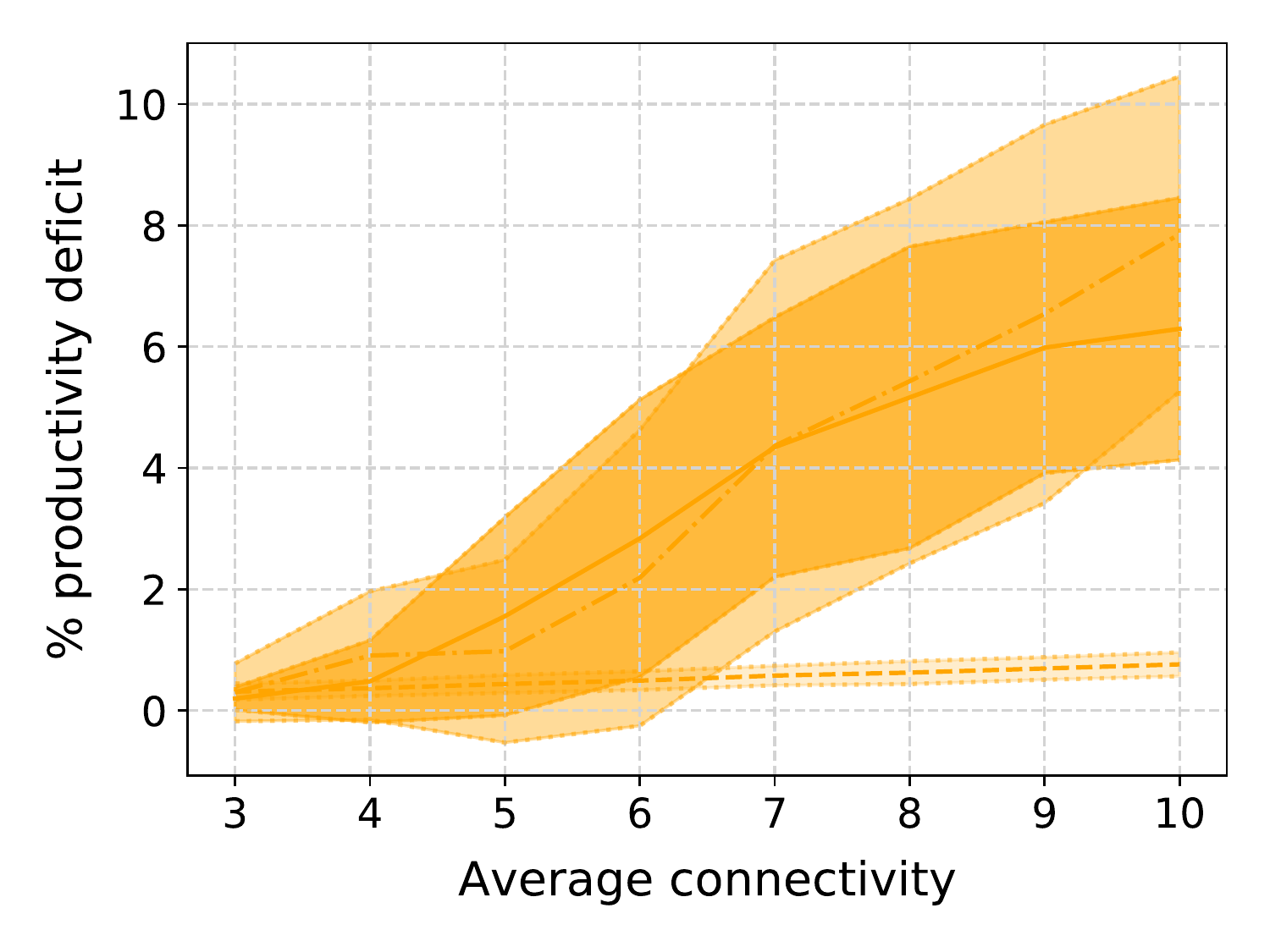}
\includegraphics[width=0.45\textwidth]{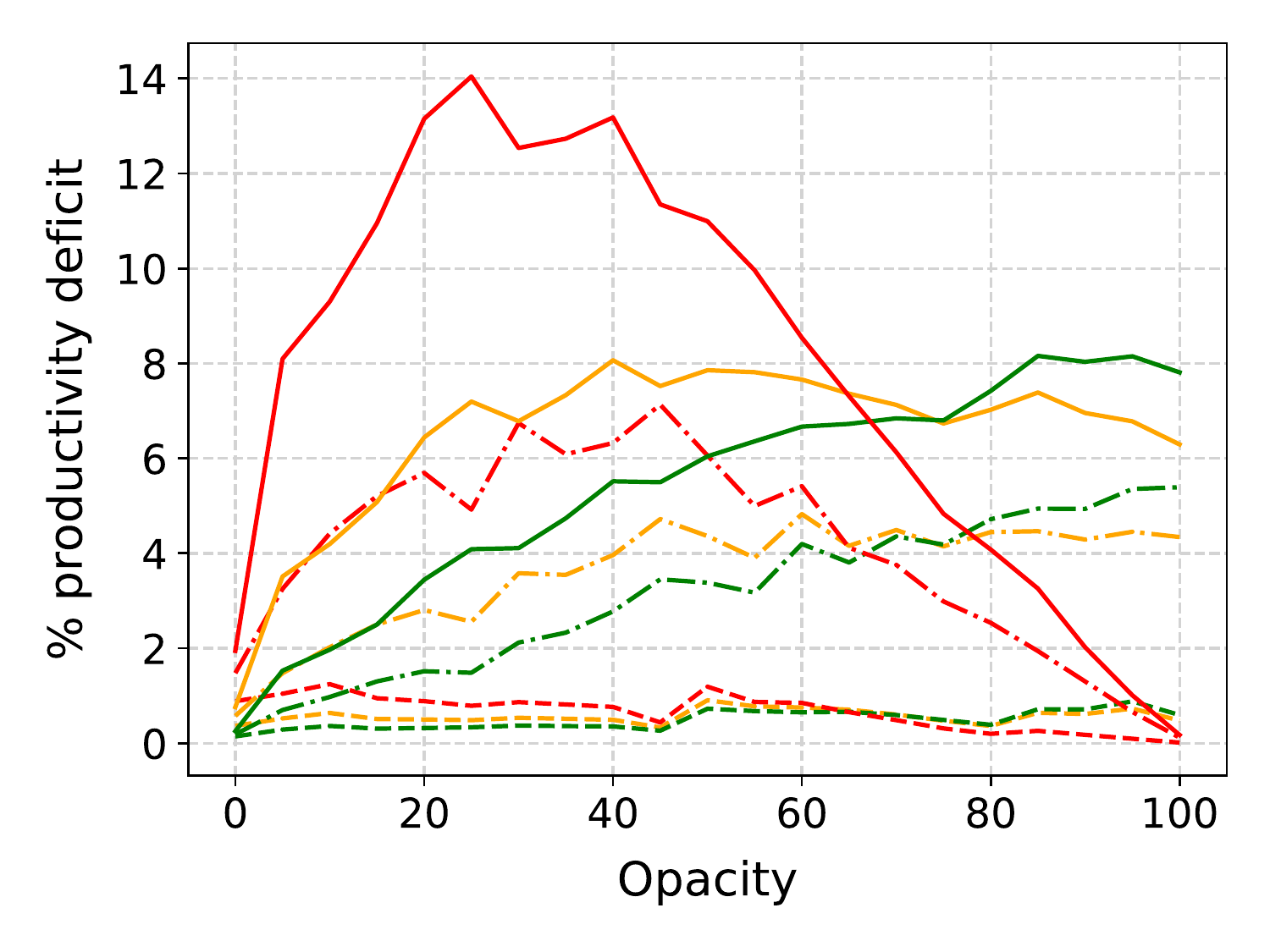}
\includegraphics[width=0.45\textwidth]{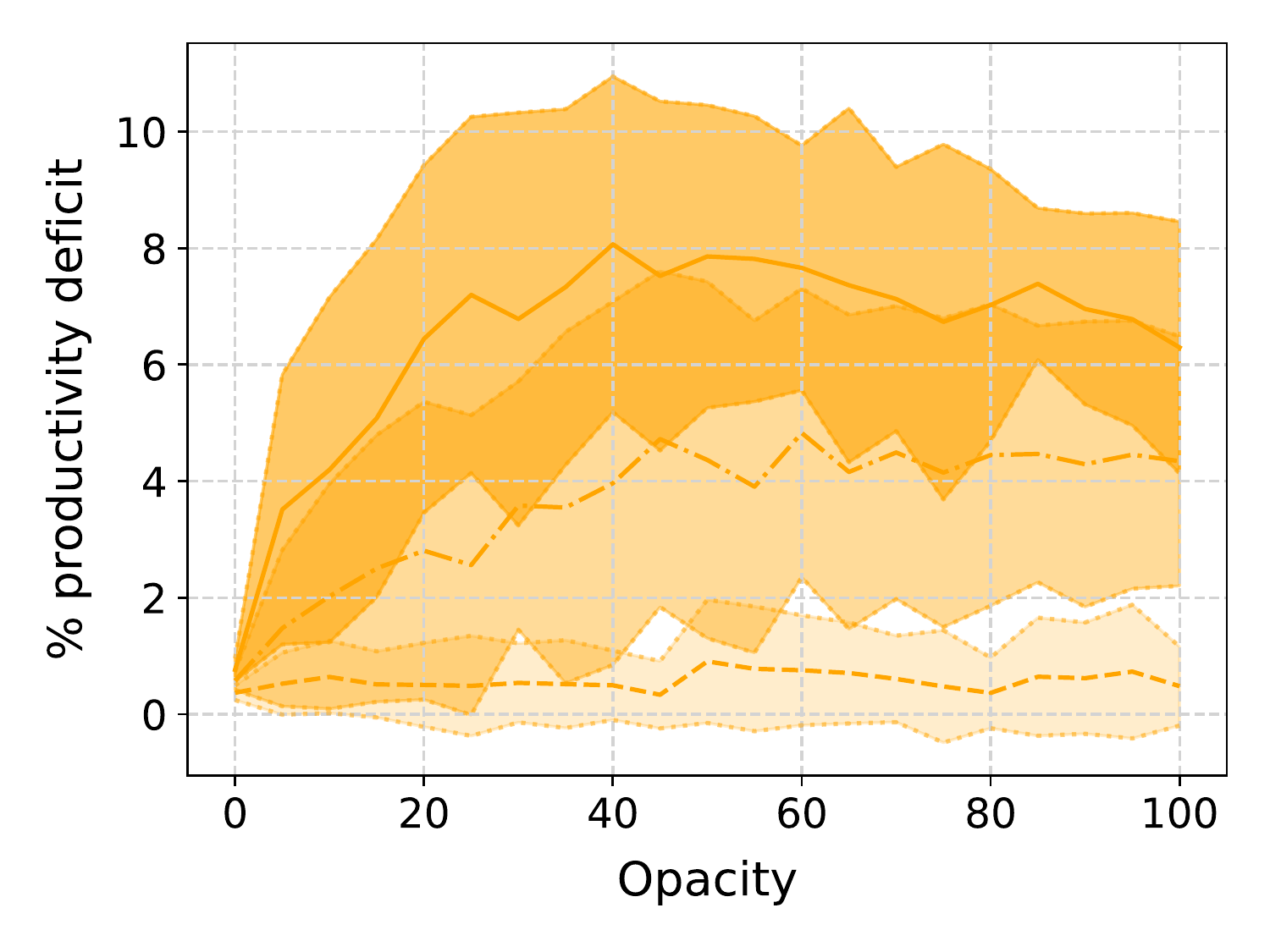}
\caption{(Top left): 
Productivity deficit as a function of degree (top figures)
and opacity (bottom figures) for the three different types of working environment --- academic (green lines), office (orange) and factory (red) --- and three different opacities --- $0\%$ (dashed lines), $50\%$ (dot-dashed) and $100\%$ (solid).  (Top Right): Mean, plus and minus one standard deviation for the office scenario. (Lower panels): Same as above, but against opacity for $d=4$ (dashed lines), $d=7$ (dot-dashed) and $d=10$ (solid).}\label{fig:cold4}
\end{center}
\end{figure}

From these simulations, we can also compute the productivity deficit according to \eqref{eq:Prod} and \eqref{eq:Pdef}. The results are presented in Fig.~\ref{fig:cold4}.
The results for the ``academic" working environment (green lines) are as expected. In this environment, well individuals are equally productive at home as in the workplace. So productivity is greatly enhanced by high transparency. 
 In the case of factories and mixed offices, especially the factory, the productivity curve is more U-shaped and it may seem that optimal productivity can be gained at 100\% opacity. But at what price? In this scenario, for the higher degrees, the majority of the workforce have caught the virus, and it seems clear that in reality the factory will need to be closed down, because it has become a hotbed of infection. 
\subsection{Running hot}

\begin{figure}
\begin{center}
\includegraphics[width=0.3\textwidth]{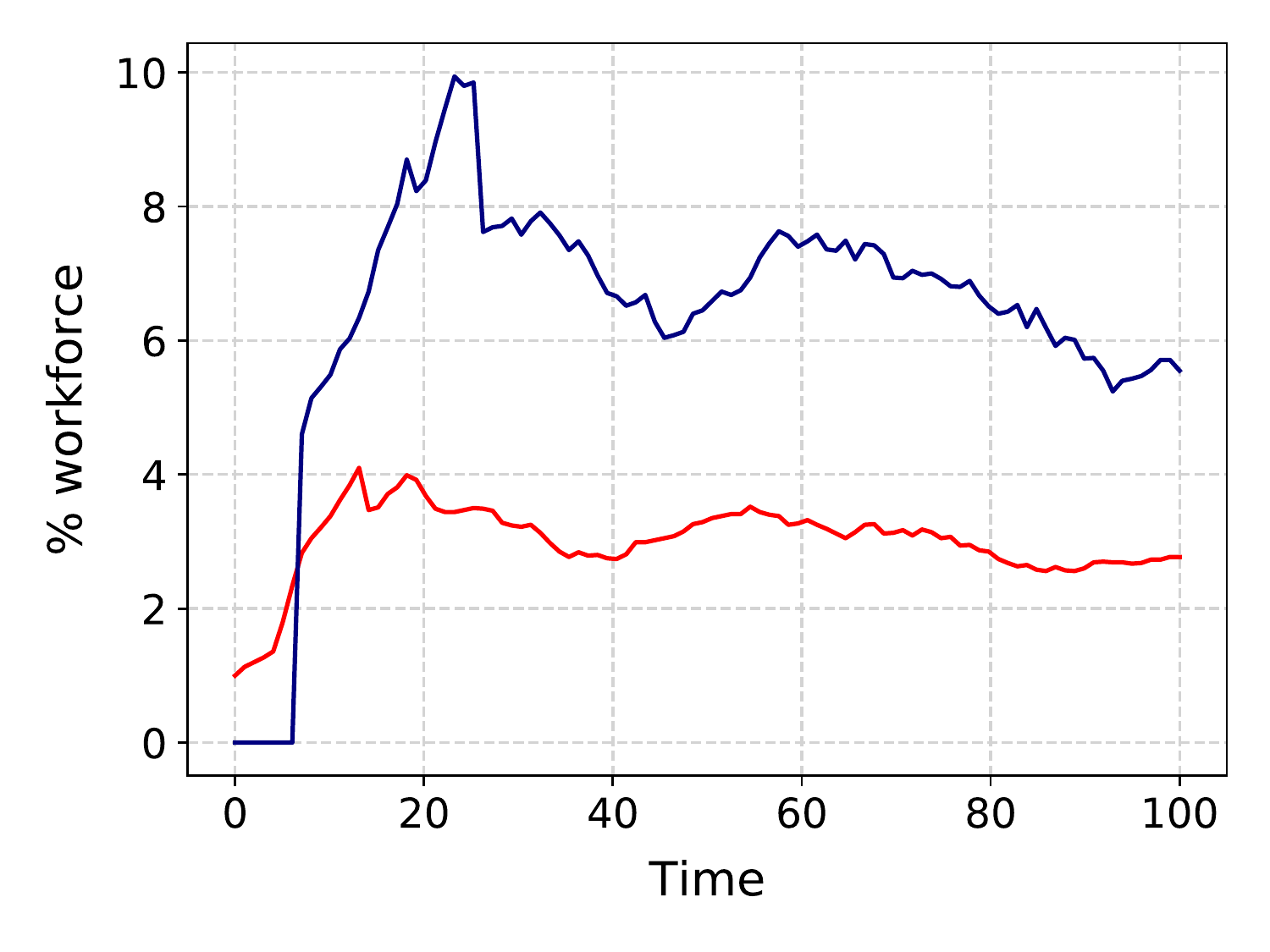}
\includegraphics[width=0.3\textwidth]{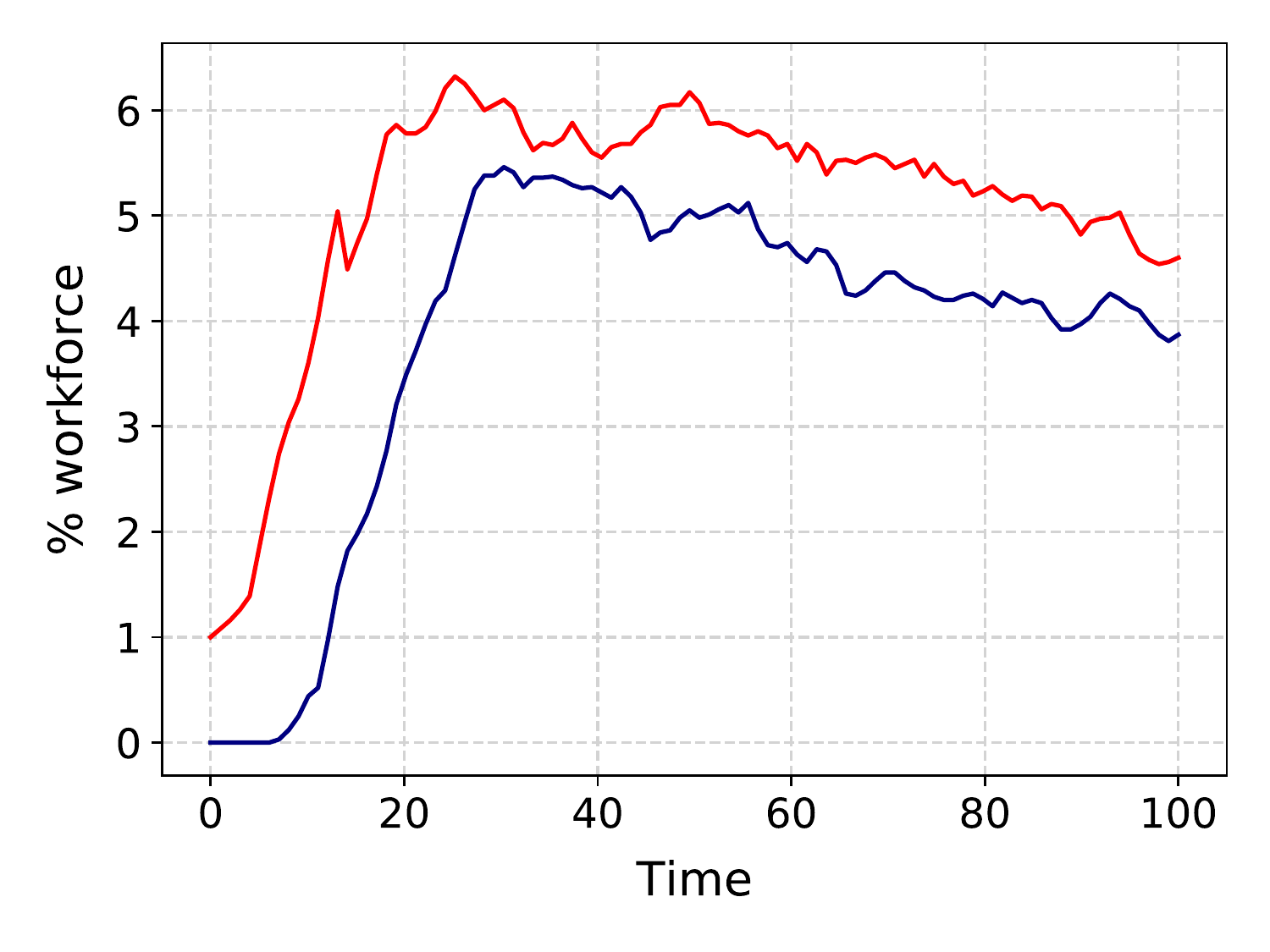}
\includegraphics[width=0.3\textwidth]{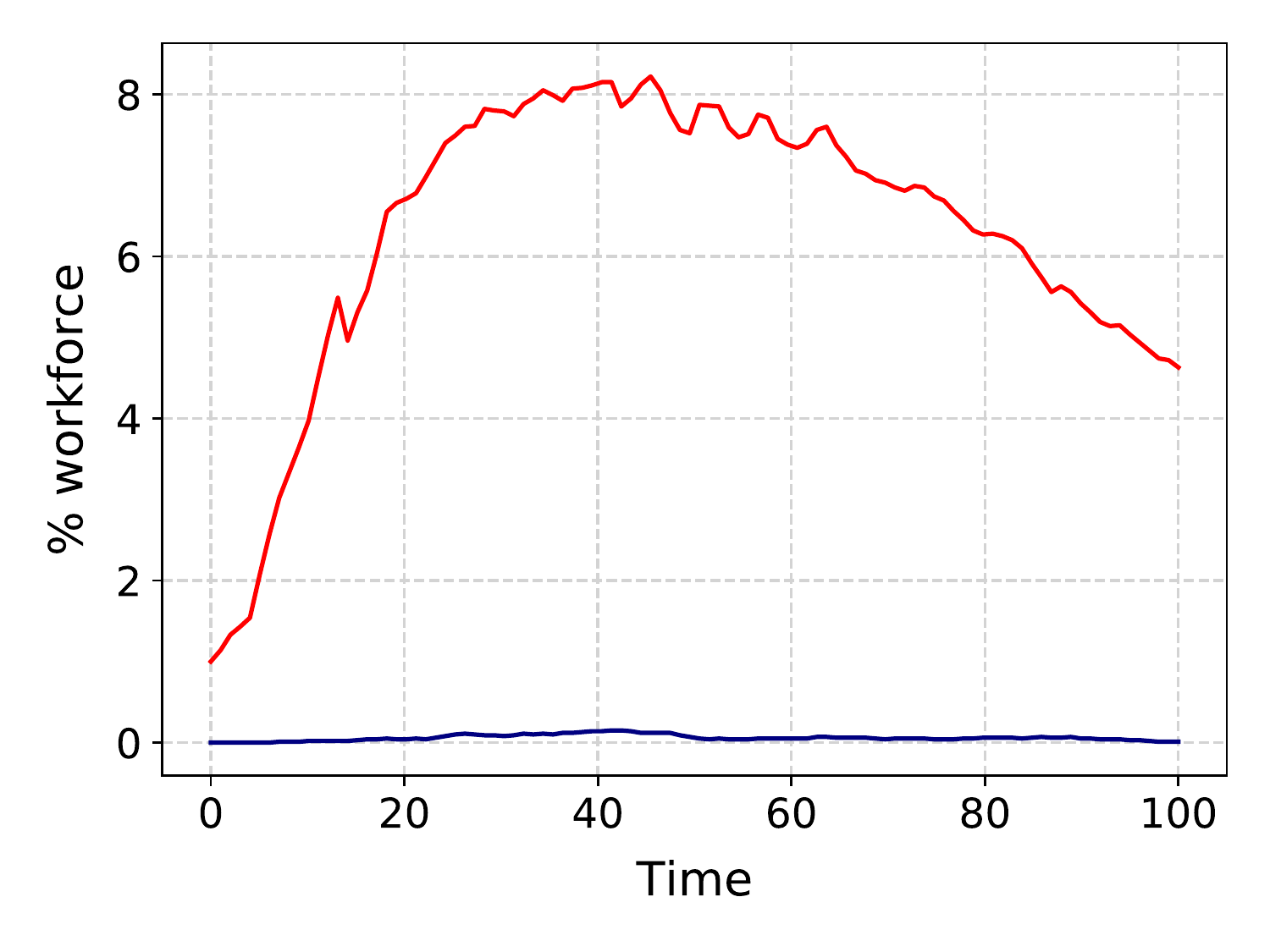}
\includegraphics[width=0.3\textwidth]{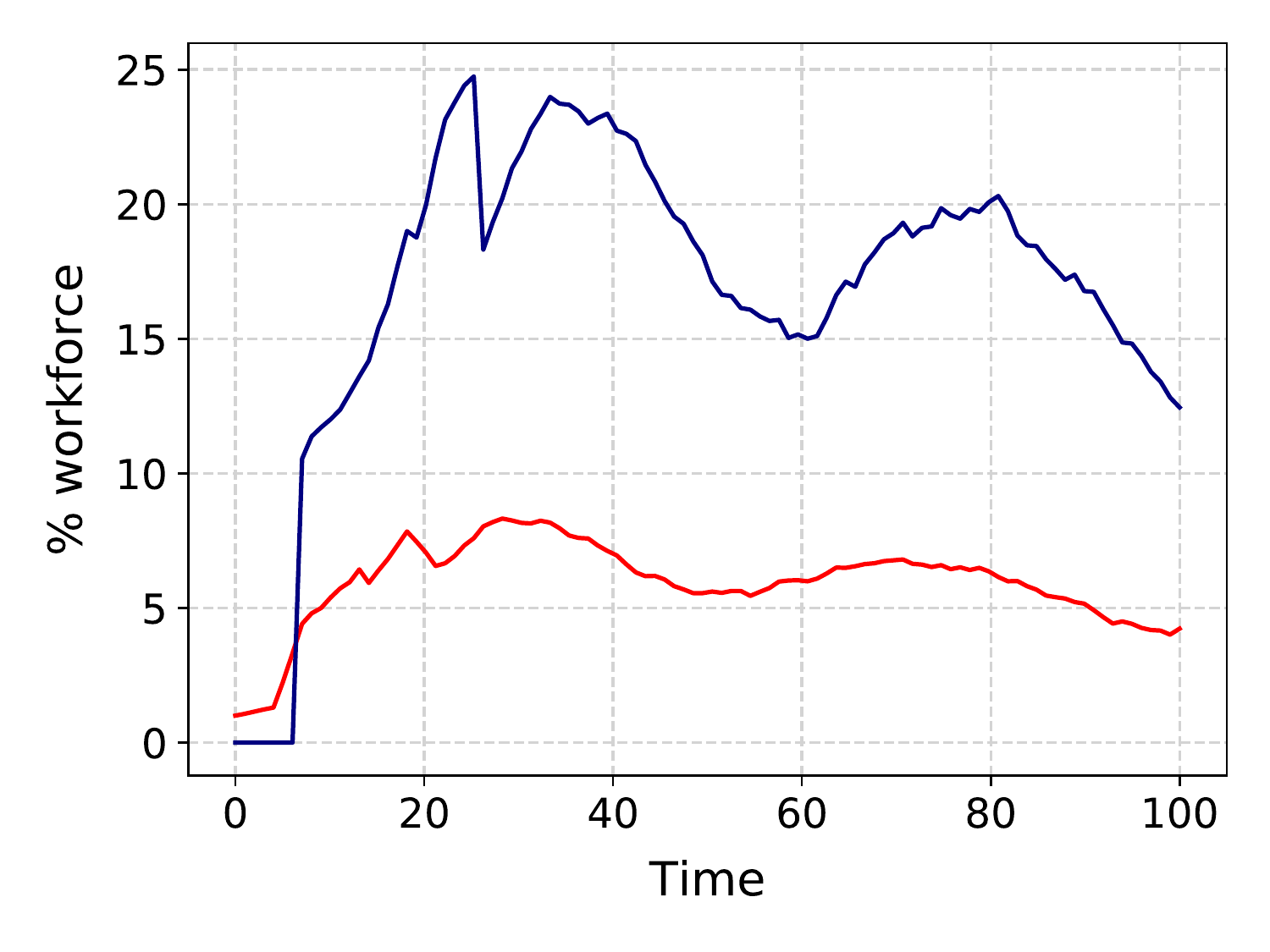}
\includegraphics[width=0.3\textwidth]{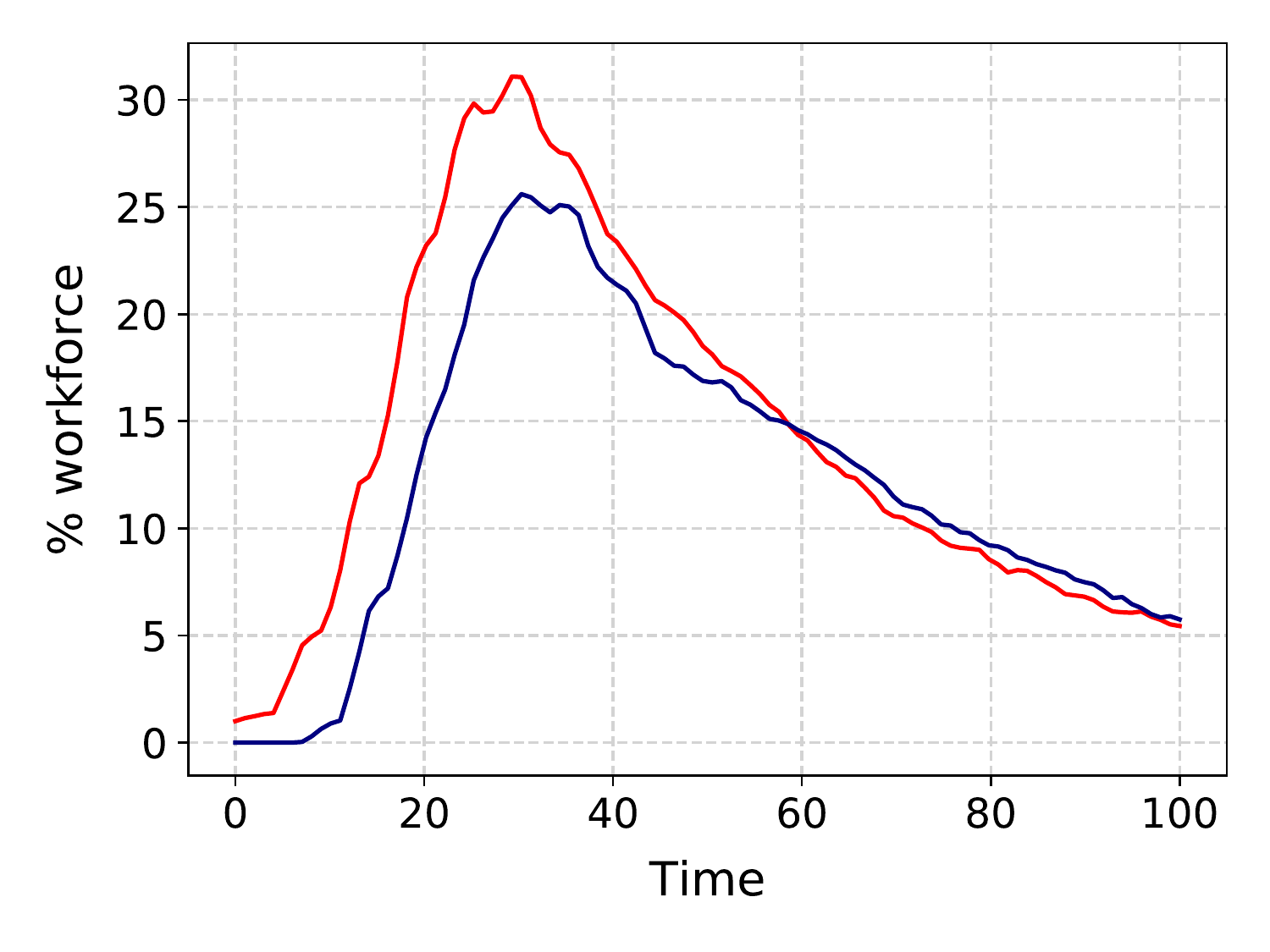}
\includegraphics[width=0.3\textwidth]{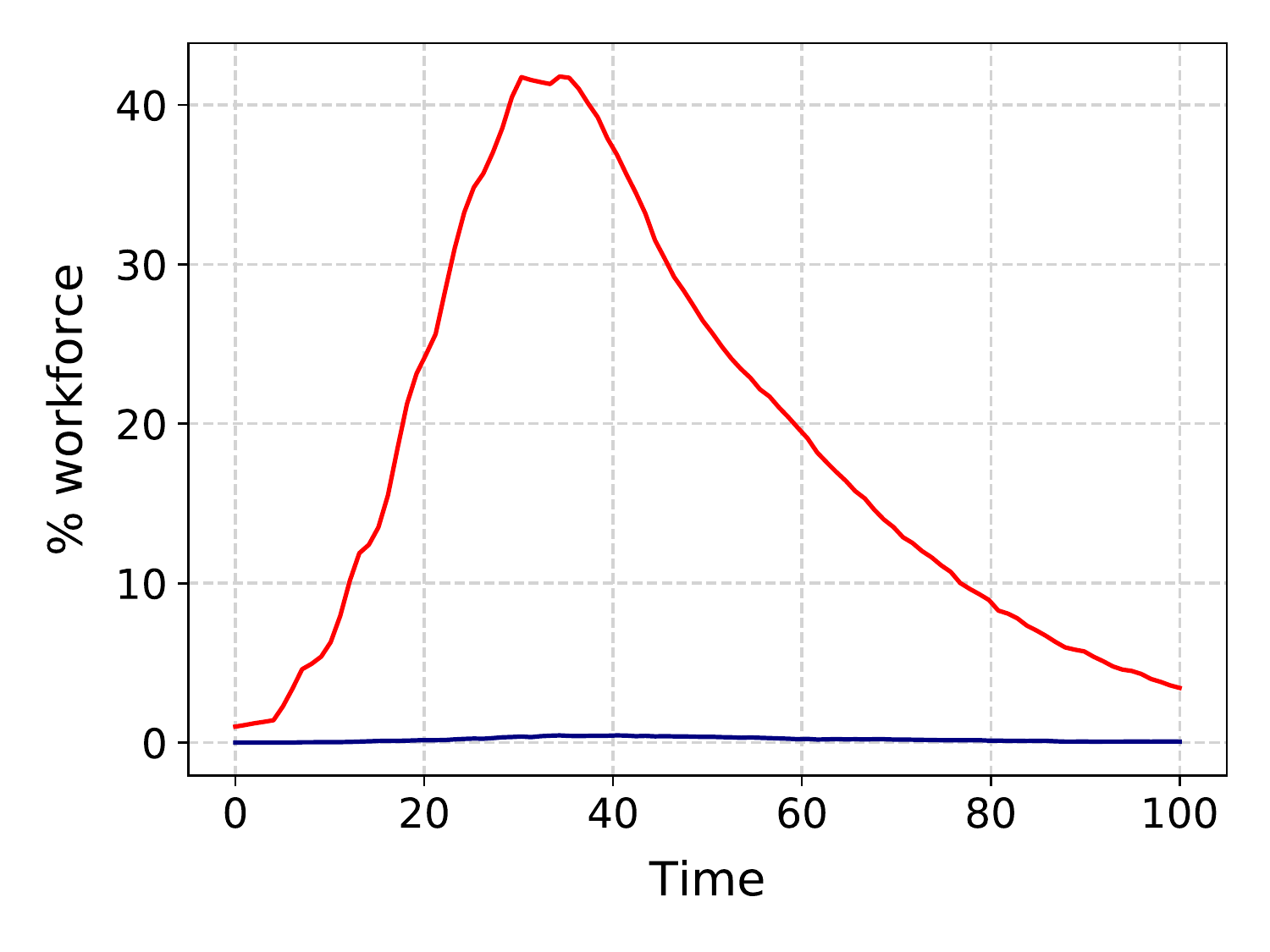}
\caption{Same as Fig.~\ref{fig:cold1} but in the case of hot runs.}
\label{fig:hot1}
\end{center}
\end{figure}

\begin{figure}
\centering
\includegraphics[width=0.45\textwidth]{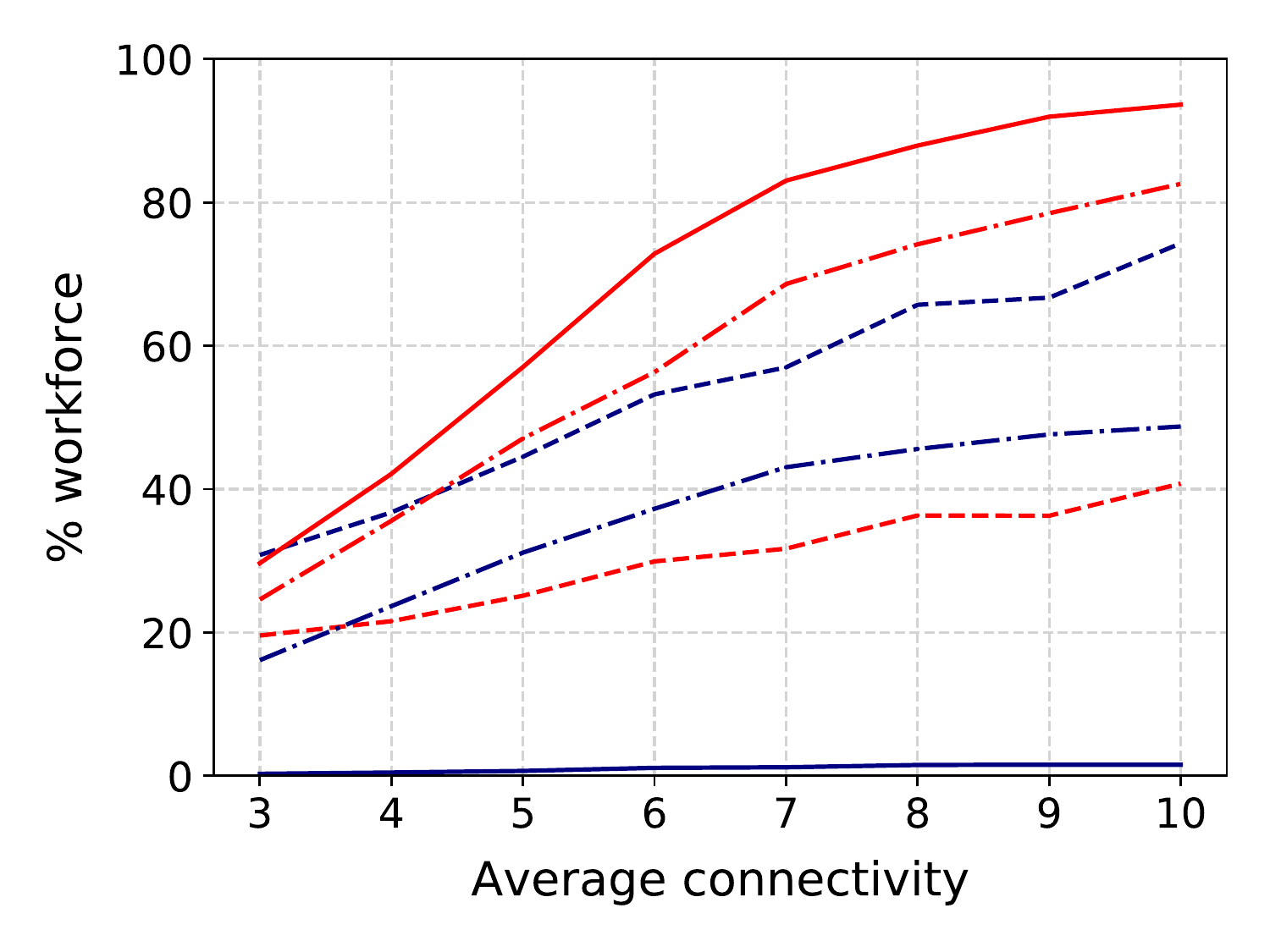}\includegraphics[width=0.45\textwidth]{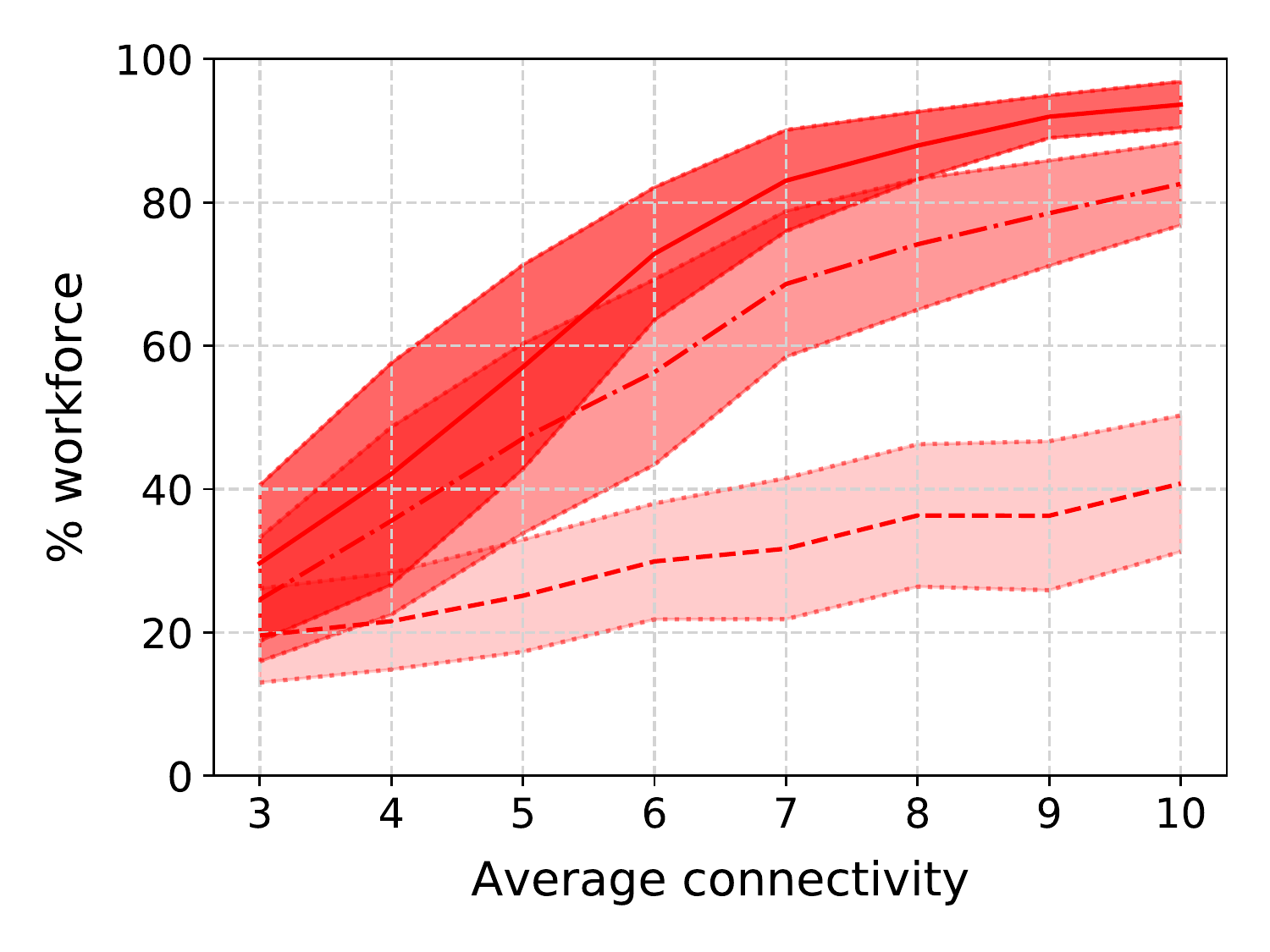}
\caption{Similar to Fig.~\ref{fig:cold2} but for running hot.}
\label{fig:hot2}
\end{figure}

\begin{figure}
\centering
\includegraphics[width=0.45\textwidth]{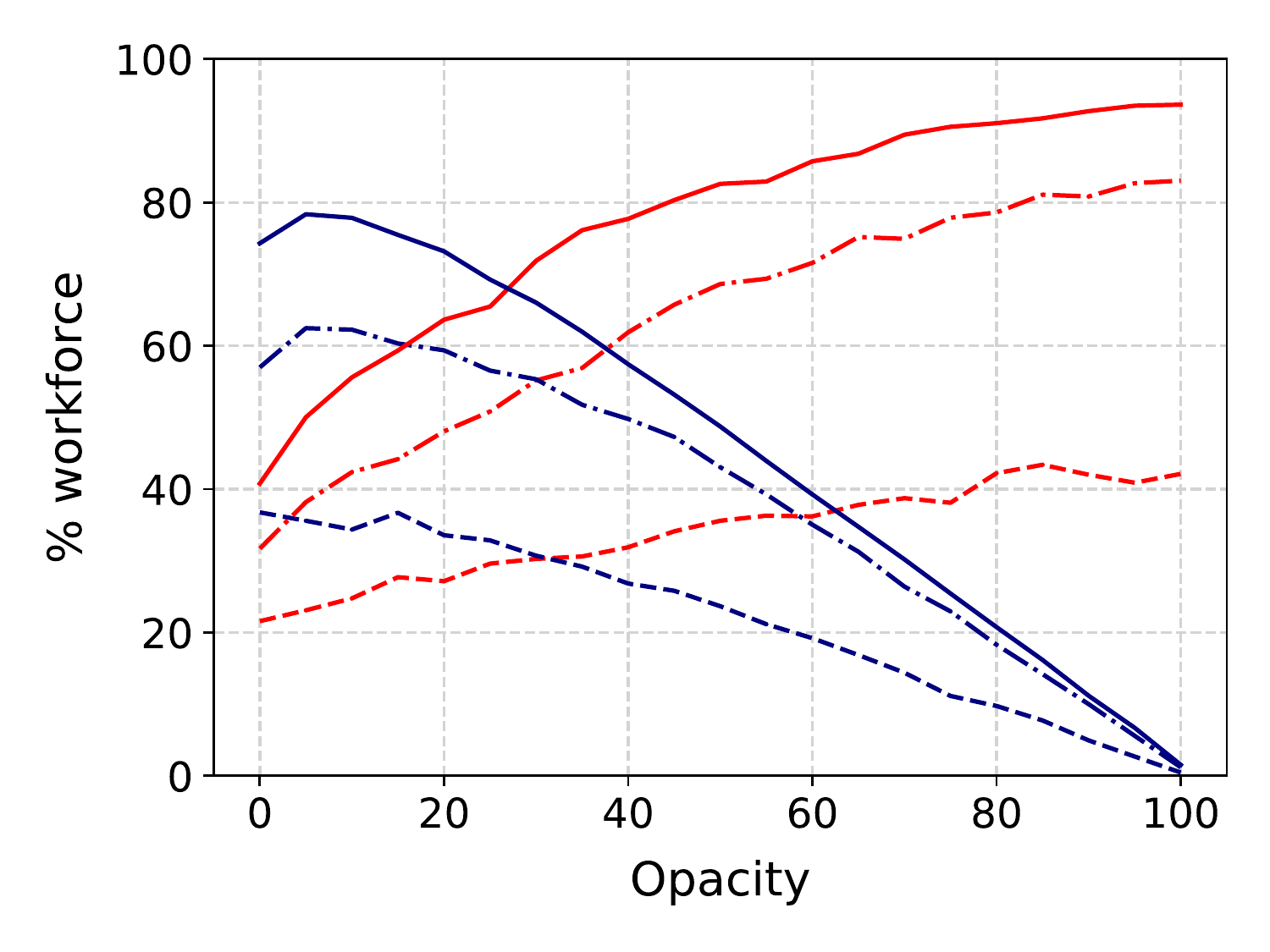}\includegraphics[width=0.45\textwidth]{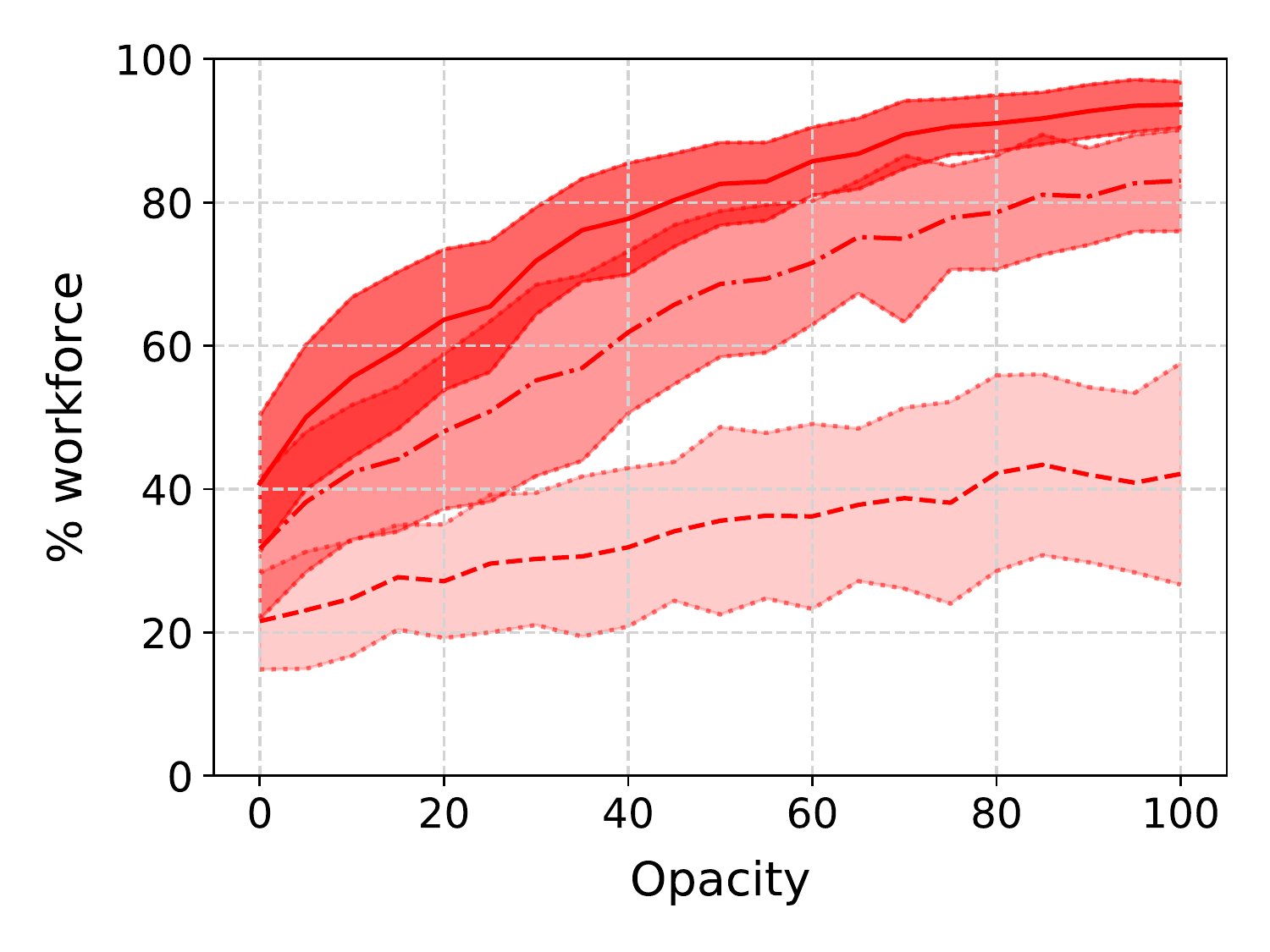}
\caption{Similar to Fig.~\ref{fig:cold3} but for running hot.}
\label{fig:hot3}
\end{figure}

\begin{figure}
\begin{center}
\includegraphics[width=0.45\textwidth]{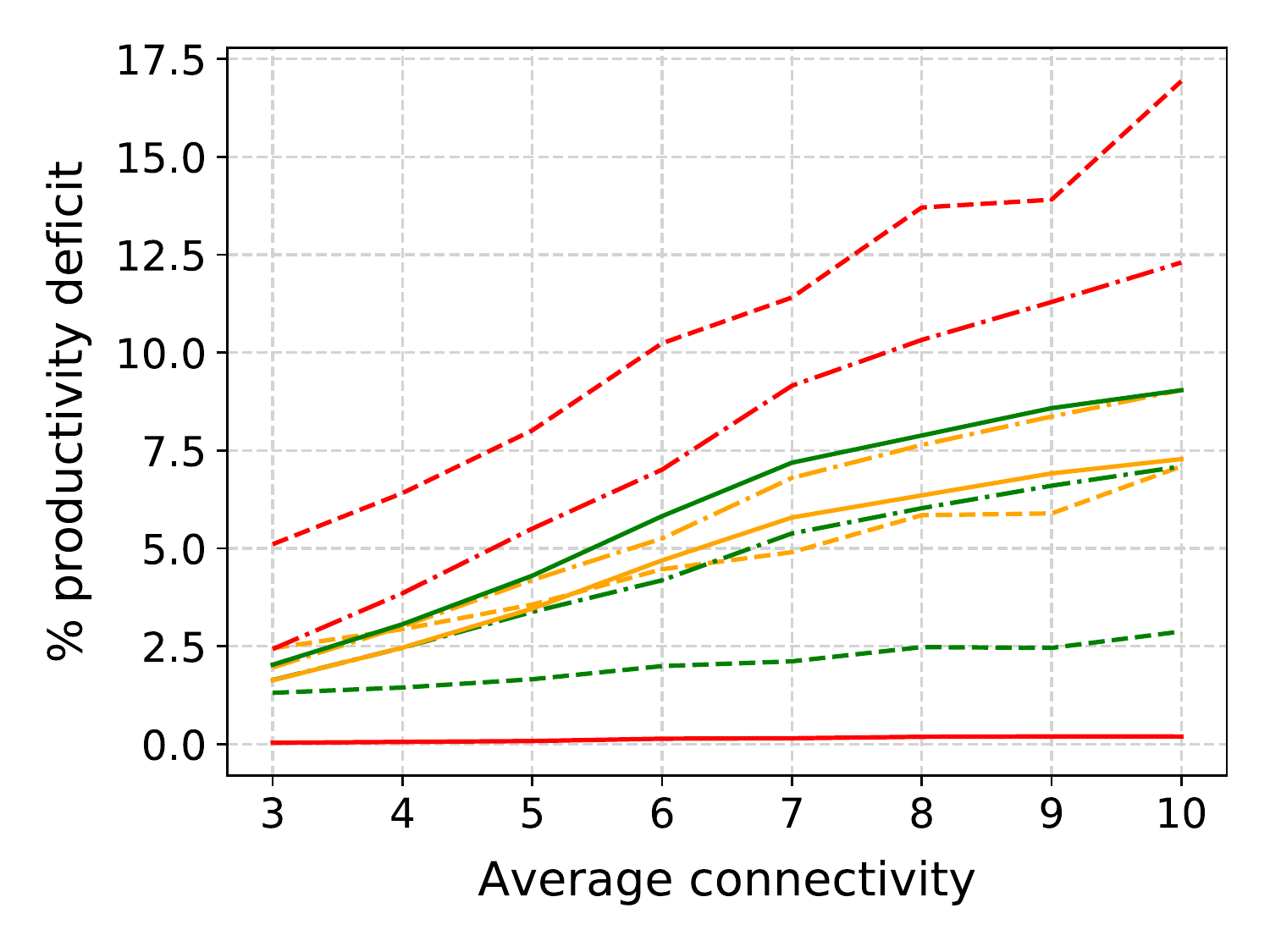}
\includegraphics[width=0.45\textwidth]{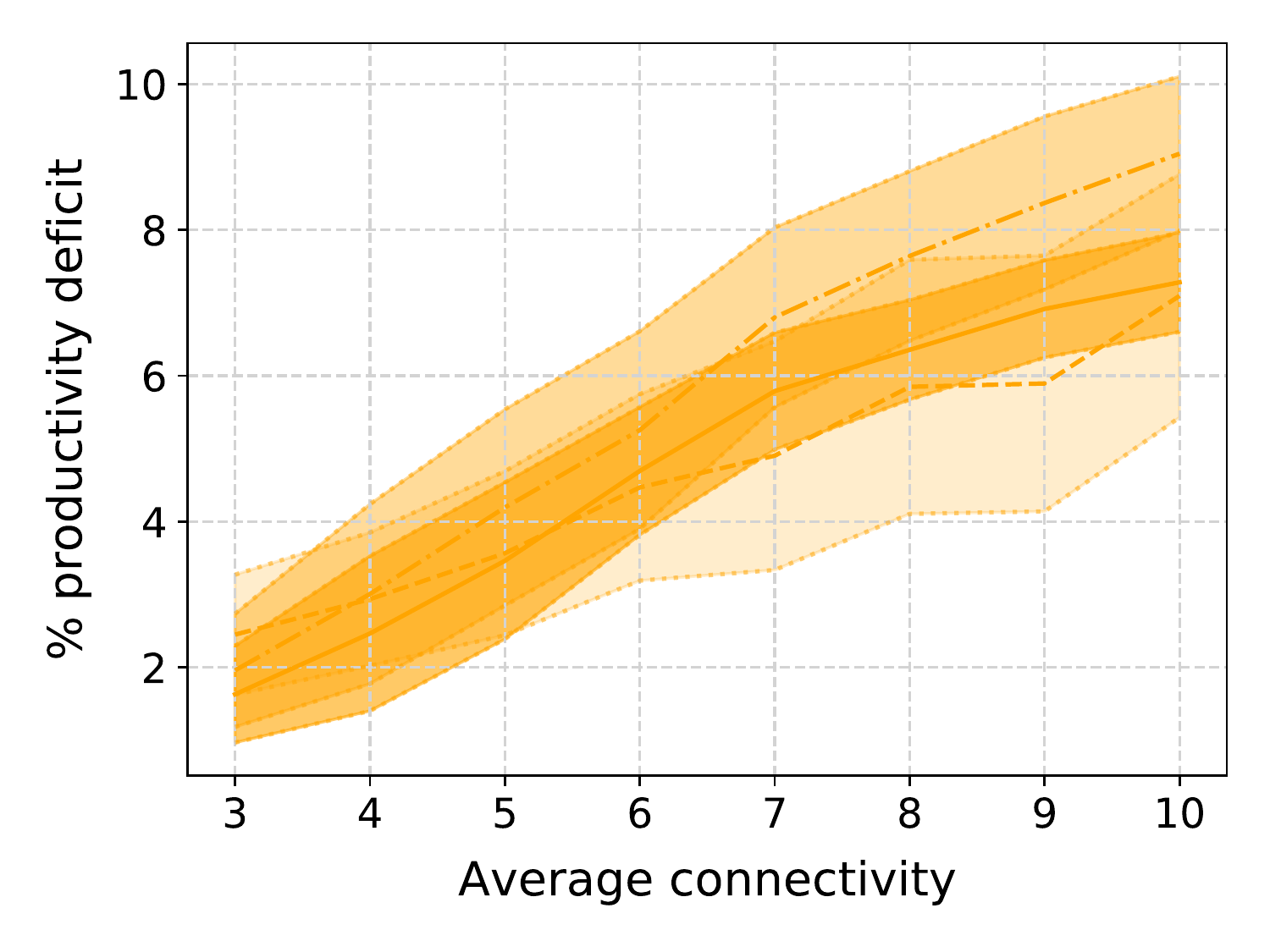}
\includegraphics[width=0.45\textwidth]{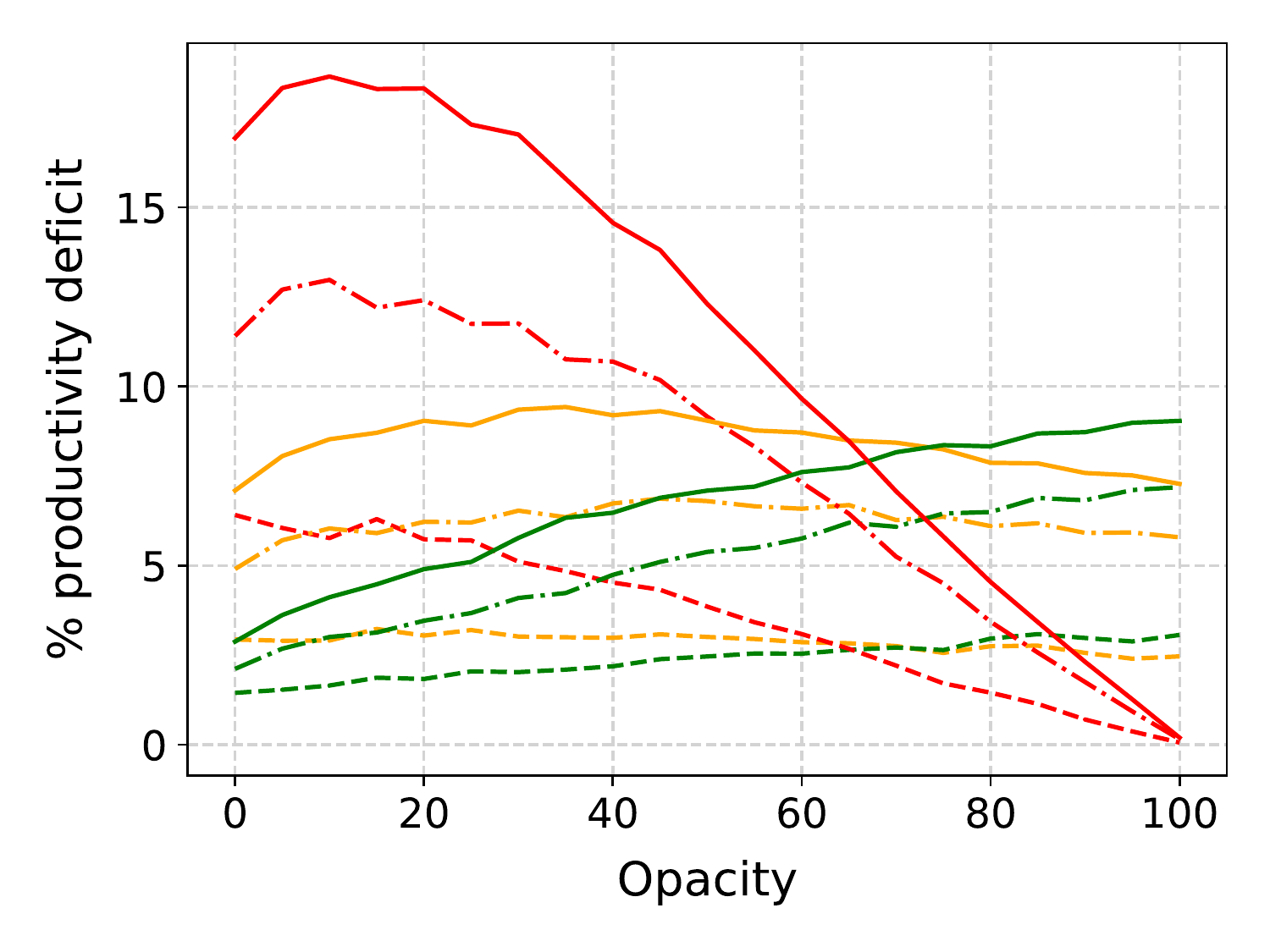}
\includegraphics[width=0.45\textwidth]{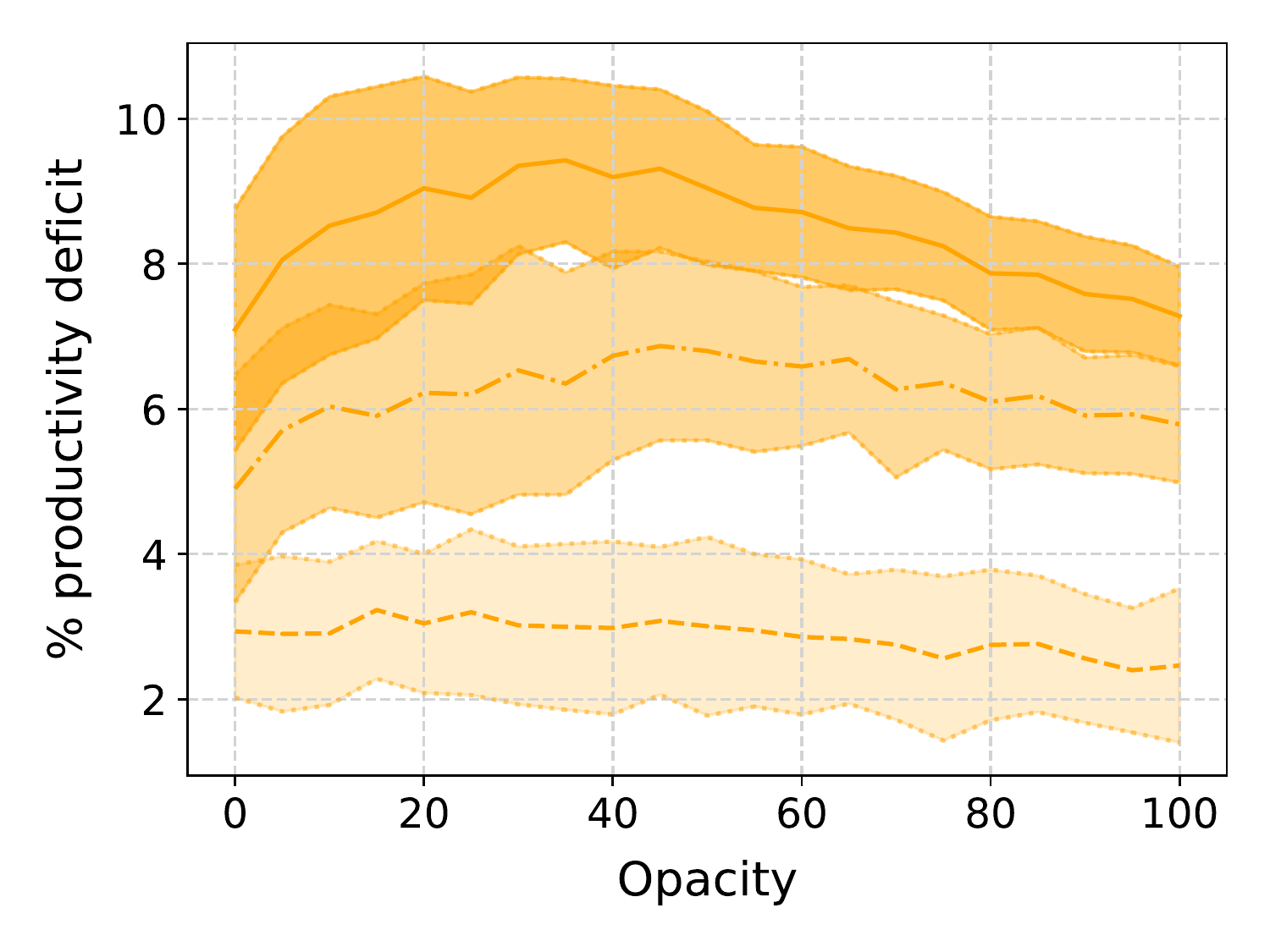}
\caption{Same as Fig.~\ref{fig:cold4} but for running hot}\label{fig:hot4}
\end{center}
\end{figure}

We have performed exactly the same set of simulations (with fresh randomisations, of course) in the case of running hot (cf.~(\ref{eq:hunt_the_thimble})).
Here the same initial condition is used as for the cold case,  with one randomly chosen exposed individual (state $E$) at $t=0$ with all others in state $S$. The difference now is that, with 
$N=100$ and $\alpha=0.001$, approximately every 10 days a new infected individual is likely to enter the workplace. 
Data that are exactly analogous to those in the cold case are presented in Figs.~\ref{fig:hot1}--\ref{fig:hot4}.

It is worth commenting on the different time-course dynamics in Fig.~\ref{fig:hot1} than Fig.\ref{fig:cold1}. In the case of a completely transparent workforce (left hand panels) note how the infection does not die out for the case of running hot. This is because approximately every 10 days a new infection enters the workplace. Thus the number of sick individuals (about 3 for $d=4$ and about 7 for $d=10$) remains constant as does the number of workers at home (about 8 and 20 respectively) throughout the simulation. Note that the number who are sick at any one time is roughly $d$ times the number who would be expected to be sick if there was no contact within the workplace.  
For either 
$50\%$ or $100\%$ opacity, in the case of low degree (upper middle and right panels), the running hot case causes the infection to last much longer in the workplace, with a much wider peak than the running cold case. The maximum number of infected individuals also rises, compared to the cold runs, which is even more apparent in the case of of $100\%$ opacity. 

These observations, based on just isolated simulation runs, are born out in general, in the Monte Carlo parametric runs in Figs.~\ref{fig:hot2}--\ref{fig:hot4} which, apart from the value of $\alpha$, are run under exactly the same assumptions as Figs~\ref{fig:cold1}--\ref{fig:cold2} in the cold case.  The results here are broadly similar to the cold case, but note the dire consequences of large opacity in terms of the spread of the disease. 

\subsection{Analysis}

A simple analysis can estimate the reproduction number 
$r_0$ in the case of different opacities and average network degree, along with the size of the infection within the workplace as a function of parameters by passing to a continuum limit. 
Taking the limit of a sufficiently long time and
sufficient network size $T,N\gg 1$, and assuming that $\epsilon=\zeta=0$ for simplicity, we obtain
the SIR-like model
\begin{align}
\dot{S}&  = \frac{(1-\delta)}{\tau} I  - \omega \frac{d}{N} \beta S I - \alpha S, \label{eq:SIR1} \\
\dot{I} & = \alpha S + \omega \frac{d}{N} \beta SI  -\frac{1}{\tau} I,  \label{eq:SIR2} \\
\dot{R}& =\frac{\delta}{\tau} I, \label{eq:SIR3}
\end{align}
where $S$, $I$ and $R$ now represent the numbers of the total workforce in states $S$, $\{I,U\}$ and
$R$ respectively as a function of time (dot represents differentiation with respect to time). Here 
$$
\tau=t_A+t_U \quad \mbox{ and } \omega = O/100\%
$$
represent the time of infection and proportion of
opaque individuals respectively,
$\delta$ is the proportion of people who gain immunity upon recovery and $d/N$ represents the chance that two individuals are connected within the workplace (the average degree of the network divided by the total number of individuals). 
Note that total population is conserved, that is $S+I+R=N$ is constant.

We first nondimensionalise the model (\ref{eq:SIR1})--(\ref{eq:SIR3}), scaling populations with the total population $N$ and time with the recovery timescale $\tau$ such that
\begin{eqnarray}
 (S, I, R) = N(S^*, I^*, R^*), \qquad t=t^*/\tau,
\end{eqnarray}
to find
\begin{align}
\dot{S^*}&  = (1-\delta) I^*  - r_0 S^* I^* - a S^*, \label{eq:SIR1non} \\
\dot{I^*} & = a S^* + r_0  S^*I^*  -I^*,  \label{eq:SIR2non} \\
\dot{R^*}& =\delta I^*, \label{eq:SIR3non}
\end{align}
where 
\begin{equation}
r_0=\omega d \beta \tau \quad \mbox{ and }
a=\alpha \tau. 
\end{equation}
Note that $r_0$ is the 
the basic {\bf reproduction number} of the outbreak, representing the average number of new infections one infected person generates within the workplace.  The parameter $a=\alpha \tau$ is the balance of timescales between new infections entering the workplace and infected individuals recovering (recall $a=0$ is the ``running cold'' scenario such that no new infections are entering the workplace). Along with the immunity rate $\delta$, which is already dimensionless, these three parameter groupings control the behaviour of the infection within the workplace and will determine the dynamics of the disease. 
Total population is still conserved, now scaled such that $S^*+I^*+R^*=1$.

\begin{figure}
\centering
\includegraphics[width=0.45\textwidth]{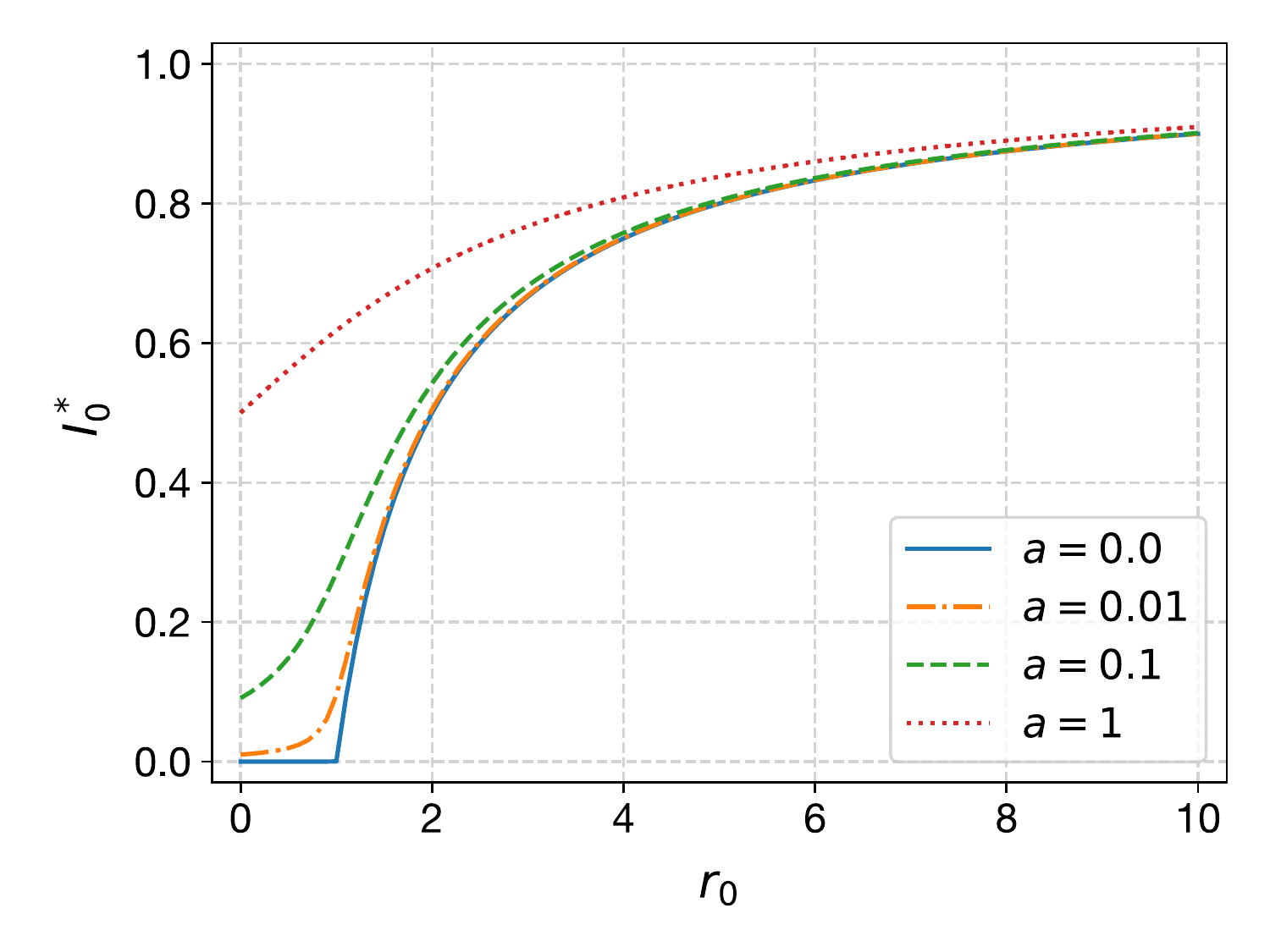}
\includegraphics[width=0.45\textwidth]{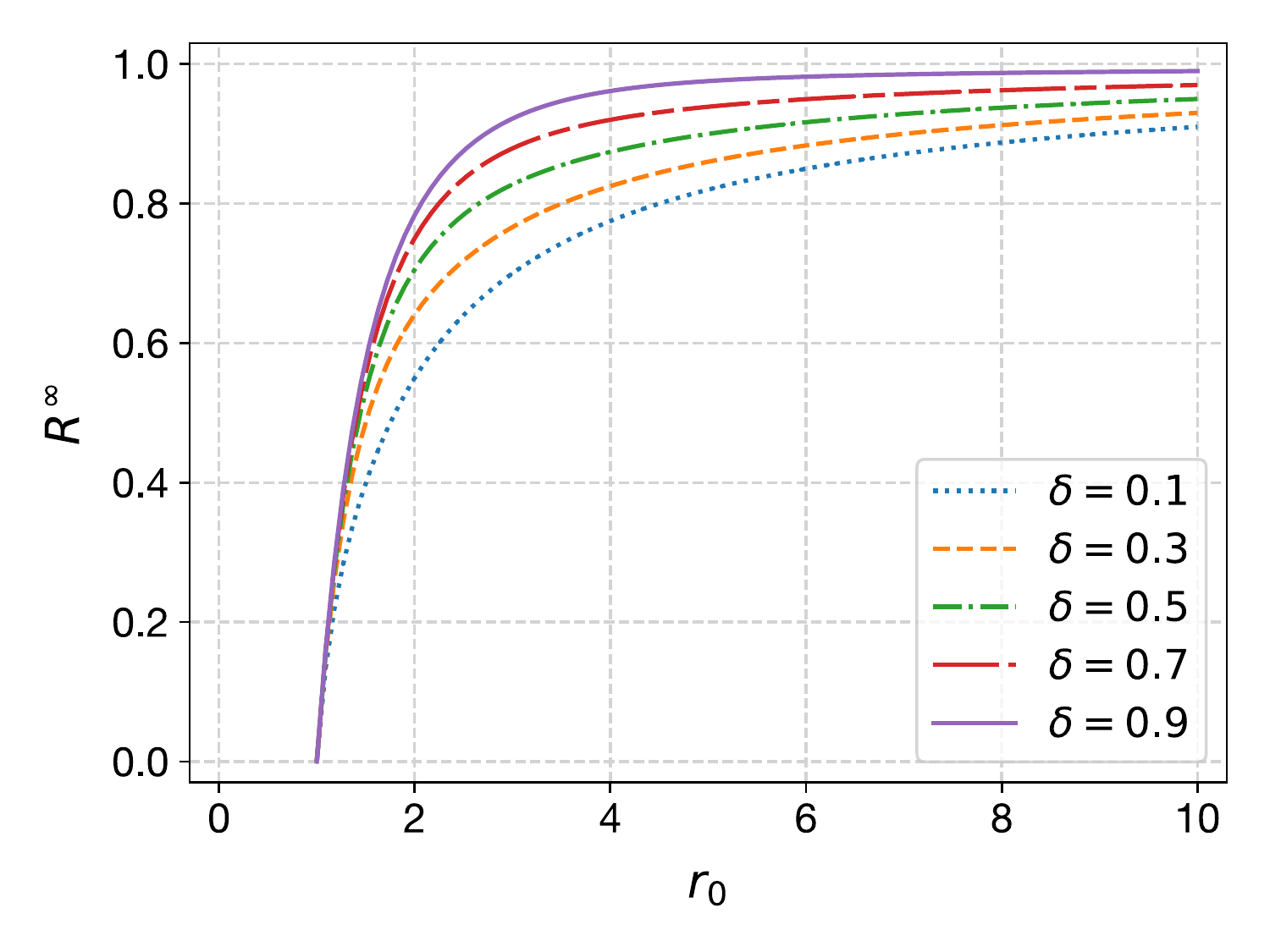}
\caption{(Left) Plot of $I_0^*$ given by \eqref{eq:I0*} against $r_0$ for different $a \geq 0$ in the case of no immunity ($\delta=0$). (Right) Positive value of $R^\infty$ obtained
by solving \eqref{eq:Rinfty} for different values of $\delta>0$ in the running cold case ($a=0$).}
\label{fig:analytic}
\end{figure}

The immediate benefit of non-dimensionalisation is that the relative importance of the parameters
of the simulation becomes obvious.  For example 
$r_0$ depends on the product of $d$ and $\omega$,
which implies the opacity that a workplace can tolerate before $r_0>1$ is inversely proportional to $d$ the average number of interactions per worker. This duality between $O$ and $d$ is also apparent in the simulation results (cf. the red curves in Fig.~\ref{fig:cold2} with \ref{fig:cold3}, and Fig.~\ref{fig:hot2}
with \ref{fig:hot3}).

Moreover, steady states of the system (\ref{eq:SIR1non})--(\ref{eq:SIR3non}) indicate what we expect to happen for long times.  This varies depending on the values of parameters $r_0$, $a$ and $\delta$.

When $a>0$ is non-zero (the running hot scenario) with some level of immunity $\delta\neq 0$, the steady state  is $S_0^*=I_0^*=0$, $R_0^*=1$, that is everyone eventually catches and recovers from the disease.    When recovery confers no immunity ($\delta=0$), we find the steady state
\begin{eqnarray}
 S_0^* &=& 1-I_0^* \nonumber \\
 &=&\dfrac{r_0+1+a-\sqrt{\left(r_0-1-a\right)^2 + 4 ar_0}}{2r_0}, \label{eq:S0*}\\
 I_0^* &=&\dfrac{r_0-1-a+\sqrt{\left(r_0-1-a\right)^2 + 4 ar_0}}{2r_0}. \label{eq:I0*}
\end{eqnarray}
In this case the infection is always present within the workforce with
a level dependent on both the infection dynamics on the network
($r_0$) and the rate of infection outside the workplace $a$. (Note
that this collapses to $I_0^*=1-1/r_0$ when $a=0$ as would be
expected, showing that the infection will die out if $r_0<1$). Graphs
of $I_0^*$ versus $r_0$ are shown in the left-hand panel of
Fig.~\ref{fig:analytic}.

For the running cold scenario, such that $a=0$, with some level of immunity $\delta> 0$,  the nontrivial steady state is given by
\begin{equation*}
S^*_0=1-R^\infty, \qquad
I^*_0=0, \qquad R_0^*=R^\infty,
\label{eq:steady}
\end{equation*}
where $R^\infty$ satisfies the transcendental equation
\begin{equation} 
r_0(R^\infty-1)-\delta + 1=  \left(1-r_0-\delta\right)e^{-r_0R^\infty/\delta}.
\label{eq:Rinfty}
\end{equation}
Note that $R^\infty=0$ is always a solution to \eqref{eq:Rinfty} for
all $r_0,\delta>0$, but there are also non-trivial solutions, which we
plot in Fig.~\ref{fig:analytic} as a function of $r_0$, for various
$0<\delta<1$.  There is a (transcritical) bifurcation at $r_0=1$ such
that for $r_0<1$ the non-zero steady-state value of $R^*_0$ is
$R_0^*=0$, implying that the infection dies out. For $r_0>1$, the
non-trivial value of $R^\infty>0$ becomes the stable steady
state. Note how the $R^\infty$ curve rises steeply with $r_0$, thus
explaining the shape of infection curves in Figs.~\ref{fig:cold2} and
\ref{fig:cold3}.

\section{Discussion}
\label{sec:4}

\subsection{Observations from the simulations}

Although we have only run the simulations for a fixed set of
parameters, the analytical results suggest a certain universality to
our findings.  In particular there is an inverse proportionality
between opacity and average contact degree (since $r_0=\omega d \beta
\tau$); the greater the average connectivity within the office, the
smaller the opacity must be to avoid the infection taking hold and
eventually reaching the equilibrium.

In order to maximise productivity, Figures \ref{fig:cold4} and
\ref{fig:hot4} also show that, in most types of workplace, it is
advantageous to have almost complete transparency. This is despite the
fact that this involves sending every contact home as soon as it is
known that they are infected. The exception is the case that we have
called a ``factory", which is the extreme case where there is no work
done from home and productivity is simply the same as attendance.
Here, of course, it is best to send no one home, so that optimal
productivity is obtained at an opacity of $100\%$. However, we have
shown that this case leads to the fastest possible spread of infection
and the largest infected population within the workplace.  In reality,
if contact tracing is being conducted throughout the general
population, such a scenario is likely to lead to the workplace being
identified as a hotbed of infection, leading to the ``factory" being
forced to shut down, negative publicity for the employer and possible
prosecution. Given the characteristic `n'-shaped nature of the
productivity deficit versus opacity curve, especially in the case of
running cold (Fig.~\ref{fig:cold4}), taking such catastrophic loss of
productivity into account, it would seem then that an optimal strategy
to maximise productivity, would be again to try to maximise
transparency within the workforce.

It might be useful to reflect on what the variable ``opacity" really
represents. The opacity of the workforce could be construed as the
availability of regular testing to that workforce.  We have said that
transparent workers go home when they enter state $U$, that is they
first develop symptoms. Instead, we could alternatively consider the
state $U$ to be the return of a positive test. Then $t_A$ represents
the length of time between catching the virus and testing positive.
In such a scenario, rather than thinking of workers as having either
helpful or unhelpful behaviours, we can consider transparency as
degree to which a rapid, accurate, regular testing regime is
undertaken in the workplace.

\subsection{Policy implications and incentives}

These results suggest that making a workplace safe to reopen in the
post peak phase of a pandemic such as COVID-19 requires the adoption
of a number of changes to the running of the workplace and new
behaviours on the part of both employers and employees. Employees will
have to, for example, declare to their employers when they have
developed symptoms of the virus or tested positive, to identify those
in the workplace with whom they have been in contact and to quarantine
at home for 14 days. Employers will have to facilitate these
communications and put in place arrangements to back-fill posts left
empty through sickness absence and quarantine. In a running cold
scenario, where infection in the general population is sparse, there
could additionally be a requirement for employees to report when they
have been exposed to the virus outside of the workplace; to keep the
workplace safe, workplace contacts of this employee could also be
required to quarantine

There are well-defined and evidence-based behavioural science
principles that can be used to inform how to support and encourage
required changes in behaviour. These have been summarised as they
apply to reducing infections during the pandemic in a number of recent
reviews and commentaries
\cite{bonell2020harnessing,chater2020behavioural,west2020applying} and
have obvious application to plans to make return to work safe. They
suggest that to be effective, a workplace campaign would need to:
\begin{enumerate}
\item create a collective viewpoint emphasising how people can look after each other, rather than how individuals can look after themselves;
\item messages and requests for changes to behaviour need to be from
  trustworthy and credible sources delivered in relatable
  terms. Actions required need to be achievable;
\item create worry but not fear amongst employees. A degree of worry
  is required to motivate uptake of protective behaviours, but worry
  to the point of fear can be paralysing, leading to denial and
  avoidance behaviours. Messages creating worry have to be paired with
  appropriate, achievable actions that employees can take, which will
  reduce the anxiety and so reinforce the behaviour;
\item ensure that whatever employers and employees are being asked to
  do, they have the {\em capability}, {\em opportunity} and {\em
    motivation}.  All three are required for behaviour to occur. For
  example, people need the knowledge (capability) of how to wash their
  hands effectively, convenient facilities in which to wash their
  hands (opportunity), and belief that it is important for their
  safety and the safety of others that they wash their hands
  (motivation). Only then will they reliably and effectively wash
  their hands (behaviour).
\end{enumerate}
To achieve the required behaviours, it also needs to be specified and
it has to be demonstrated how behaving in this way will produce the
desired outcome, but also avoid unintended negative consequences. This
might include, for example, unintentionally increasing employees
social-isolation. This requires spending time understanding the
implications of the behaviours employers and employees are being asked
to perform.  This can only be done through investing in conversations
and consultations. Messages and instructions need also to be made
really clear and direct. Anxious people do not process complex
information so well.

All of the above should together create a clear, engaging rationale
and set of actions and new social norms for the workplace.  For some
workers, including those on zero hours contracts, compliance with new
workplace requirements may involve loss of earnings and discourage
them from being transparent with their employers. Equally, employers
may be anxious about loss of productivity and profits, making it
difficult for them to comply. This suggests that in both cases
compliance needs to be incentivised. The effectiveness of financial
incentives in encouraging changes in behaviour appears to depend on
their framing and on exactly which behaviour is being encouraged. It
is well-established that incentives offered to compensate for
potential losses may be particularly effective as motivators, implying
that the potential to lose anticipated earnings may be more likely to
motivate behaviour than the possibility of a financial windfall
\cite{vlaev2019changing}. This suggests that compensating for loss of
income for both employer and employee would be important components of
arrangements for safe return to work. The independent SAGE committee
report \cite{indSAGE} on how best to support effective application of
testing for the virus and contact tracing recommends: \emph{`Support
  for isolation should be provided: financial support to compensate
  lost income support for obtaining groceries etc; accommodation for
  those who cannot isolate in current residence; follow up to check
  symptoms and wellbeing'} This is not to deny the influence of
non-pecuniary motives on people's behaviour. Concerns about social
disapproval and fairness are likely to interact with, and have
reinforcing effects on, compliance with requirements for transparency
and quarantining \cite{fehr2002psychological}.

Compliance is also more likely if employers and employees feel a sense
of ownership and control over the way that the workplace return to
work scheme is designed and run. For employers and employees to
co-design schemes would give all involved a sense of control over the
plans and processes. A perception that one has control over aspects of
the workplace and job role have long been linked to increased sense of
well-being, lower perceived stress and better general health (see for
example \cite{steptoe2004influence}). Building a sense of ownership
over the plans for return to work are also likely to increase
compliance and reduce unintended consequences including exacerbating
inequalities. Discussions which help employers understand the impact
of transparency and quarantining on employees will allow them to
modify the scheme to lessen any negative impacts.  In the same way
that independent SAGE suggest that effective partnership with the
community is essential to successful implementation of testing and
contact tracing, so would a partnership between employer and employees
be needed in any workplace for infection rates to be minimised
\cite{indSAGE}.  Experience during the 2014-15 Ebola outbreak in
Sierra Leone suggests that allowing communities some control over how
they respond to the health threat improved their sense of well-being
and resulted in a reduced number of deaths associated with the
pandemic \cite{Richards}.

The ability to make choices about how processes are carried out may
also increase compliance. Qualitative evidence from studies of
sexually transmitted infections suggest that giving people choice over
whether they notify contacts themselves versus having someone else
undertake this task increases cooperation with contact tracing
\cite{edelman2014opportunities,kissinger2019o12}.

Engaging the whole workforce in planning safe return to work might
also improve compliance through engendering a sense of collective
efficacy; in other terms, a community spirit which will be necessary
to persuade people to be transparent about their infection status and
quarantine themselves. Recent evidence review suggests that key to
adherence to quarantine were clear understanding of the disease and
quarantine procedures, social norms and perceived benefits of
quarantine, perceived risk of the disease and practical problems such
as maintaining supplies and financial consequences of being out of
work \cite{webster2020improve}.  In light of this, Government advisory
bodies have issued guidance to encourage quarantining that includes
emphasising civic duty, advertising the changing social norms and
allowing others in the community to express disapproval and stressing
the value of the organisation, in this case, the employing
organisation.

\section{Conclusion and recommendations}
\label{sec:5}
It is helpful to make some recommendations for policy makers, both
government and employers, as well as to the behaviour of employees in
the kind of closed, fixed-interaction workplaces envisaged in this
study.
\begin{description}
\item[Organise workplaces into small, intersecting groups.]  We
  recommend that all workplaces should seek to minimise the average
  number of work contacts per worker. Other studies have suggested
  work should take place in fixed ``bubbles" that do not interact with
  each other. For most operations, strict bubbles are not
  feasible. Instead, our results show that the necessary degree of
  transparency (the rigour to which test, trace and isolate is
  required) is directly proportional to the size of the average
  workplace interaction network.
\item[The no detriment principle.] We recommend workplaces engender a
  culture where there is no perceived detriment to a worker being
  transparent. One issue can be a workplace where there are a lot of
  self-employed or `gig economy' workers who only get paid when they
  are present. Mechanisms need to be established so that all workers
  are fully remunerated if they are required to self-isolated and
  penalised if they are found to be at work while infectious. One such
  policy intervention could be the introduction of governmental
  statutory sick pay for all workers irrespective of their contractual
  status, from day one of self-isolation.
\item[The benefit of mutualism.] As in points 1-4 in the previous
  section, and the other evidence there, we need to encourage
  workplace campaigns that are effective. A sense of shared ownership
  in maintaining the workplace infection free would need to be
  inculcated through as sense of collective efficacy. This might
  involve an employee/employer partnership that runs its own workplace
  test, trace and isolate system. This may be more effective than a
  scheme administered by an outside authority where there is the
  potential for lack of trust, for example over data privacy.
\item[Financial and other disincentives for non-compliant workplaces.]
  For some businesses, there may be a temptation for the employer to
  seek short term profit over long term benefit. If infection rates
  are low in the general population, there may be unscrupulous or
  ignorant employers who try to keep everyone in work. We saw this in
  our models for the ``factory" scenario, in which theoretical maximal
  productivity could be achieved with no transparency whatsoever. It
  should be the function of a Health \& Safety Inspectorate to provide
  large penalties to deter such behaviour which would clearly not be
  in the public good.
\item[Individual solutions for individual workspaces] Many of the
  quantitative findings in this study are a result of the generic
  parameter choices made in the simulation runs of our mathematical
  model. Some of these are disease parameters, which for COVID-19
  remain unclear at the time of writing. Others related to the nature
  of the work; for example what exactly constitutes a link between two
  workers can greatly affect the probability of infection between
  individuals. We recommend that individual employers should be
  encouraged to run a playable version of the mathematical model we
  have introduced in order to test the safety of their operation. This
  may also involve ethnographic studies both to observe how
  interactions happen in reality and also to design the optimal
  psychological and policy interventions to obtain the desired
  outcome.
\end{description}

\section*{Acknowledgements}
The authors would like to thank Julia Gog, Thilo Gross, Jessica
Enright, and Mason Porter for helpful discussions. They would also
like to thank all the organisers and participants at the VKEMS study
group, from where this work originated. Additionally, the authors
thank EPSRC, for their support of industrial and applied mathematics,
in particular through the Industrially Focused Mathematical Modelling
(InFoMM) Centre for Doctoral Training (CDT) at University of Oxford.

\end{document}